\documentclass[CRPHYS,Unicode,manuscript]{cedram}
\usepackage{blindtext}
\usepackage{epsfig}
\usepackage{color}
\usepackage{amsmath}
\usepackage{amsfonts}
\usepackage{amssymb}
\usepackage{graphicx}%
\usepackage{url}
 \usepackage{mathrsfs}

\usepackage{esvect}
\usepackage[normalem]{ulem}

\usepackage{xcolor}
\usepackage{xfrac}
\usepackage{dsfont}
\usepackage{bm}
\usepackage{array}
\newcolumntype{C}[1]{>{\centering\let\newline\\\arraybackslash\hspace{0pt}}m{#1}}
\usepackage{footnote}
\usepackage{isomath}
\DeclareMathAlphabet\mathbfcal{OMS}{cmsy}{b}{n}
\usepackage{esint} 
\usepackage{isomath}

\hypersetup{colorlinks,citecolor=blue,filecolor=blue,linkcolor=blue,urlcolor=blue}

{
 \definecolor{BLACK}{gray}{0}
 \definecolor{WHITE}{gray}{1}
 \definecolor{RED}{rgb}{1,0,0}
 \definecolor{GREEN}{rgb}{0,1,0}
 \definecolor{BLUE}{rgb}{0,0,1}
 \definecolor{CYAN}{cmyk}{1,0,0,0}
 \definecolor{MAGENTA}{cmyk}{0,1,0,0}
 \definecolor{YELLOW}{cmyk}{0,0,1,0}
}

\title{Spin Response of a Magnetic Monopole and Quantum Hall Response in Topological Lattice Models through Local Invariants and Light}

\author{\firstname{Karyn} \lastname{Le Hur}\CDRorcid{0000-0002-3990-4782}\IsCorresp}
\address{CPHT, CNRS, Institut Polytechnique de Paris, Route de Saclay, 91120 Palaiseau, France}
\email[K. Le Hur]{karyn.le-hur@polytechnique.edu}

\author{\firstname{Andrea} \lastname{Baldanza}\CDRorcid{}}
\address{Dipartimento di Fisica, Universit\` a di Roma La Sapienza, Piazzale Aldo Moro 5, I-00185 Roma, Italy}
\email[A. Baldanza]{}

\begin{abstract}
Here, we elaborate on and develop the geometrical approach introduced in K. Le Hur, Physics Reports 1104 1-42 (2025) between the magnetic monopole created from a radial field, quantum physics and topological lattice models through quantum phase transitions. We introduce an effective magnetic moment for a monopole when applying an additional source field along z-direction which also mediates the quantum phase transition. We present its relation with the transverse pumped quantum Hall current. The magnetic susceptibility can be introduced as a measure of the topological invariant i.e. it remains quantized within the topological phase until the transition. We show the relation with two-dimensional topological lattice models such as a honeycomb Haldane model in real space. We develop the theory and present a numerical analysis between local invariants in momentum space introduced from Dirac points, a real space correspondence and the responses to circularly polarized light. We develop the formalism for coupled-planes materials including the possibility of quantum spin Hall effect and address a relation between the Ramanujan infinite alternating series and an interface in real space with a topological number one-half. 
\end{abstract}
  
\begin{document}
  \maketitle

\section{Introduction}

The theory of magnetic monopoles attracts attention since the works of P. Curie \cite{Curie} and P. A. M. Dirac \cite{Dirac,Polchinski}. Detecting such monopoles associated to the planetary system, the universe, is yet of great interest. 
Recently, one of us has introduced a geometrical approach showing a correspondence between classical magnetic monopoles and quantum magnetic monopoles from a formulation at the poles on the sphere \cite{KLHReview}.
The magnetic monopole is produced from a radial magnetic field. Topological properties or equivalently the magnetic topological charge are generally measured through the integral of the magnetic field or through the integral of the Berry curvature
in quantum physics on the surface area \cite{Berry}. This approach \cite{KLHReview} can be precisely rephrased into an effective cartesian metric in the vicinity of the poles in terms of a classical vector potential or in terms of the Berry gauge potential \cite{Berry} in quantum physics associated to a coherent gauge (phase) representation for the eigenstates of the spin-1/2 particle on the Bloch sphere. Equivalently, the information on the magnetic charge is transported on each side of the equator to the poles in a thin handle. Before emphasizing on the goals in this article, we remind several practical applications \cite{KLHReview}. This approach allows for a description of the quantum Hall response when building an analogy to the Newtonian approach. When applying an electric field along the polar angle an electron goes down from north to south pole with the induction of a transverse pumped current or a topological transverse pumped charge which is related to the quantum Hall conductivity and to the physics at the edges on a cylinder. It also allows us
to show how photo-electric effects can become topological \cite{KarynLight}. The responses to circularly polarized light can measure the topological charge or the quantum Hall conductivity from the poles which can be directly applied in two-dimensional (2D) topological lattice models through the Haldane model \cite{Haldane} and the quantum anomalous Hall effect (QAH) \cite{QiZhang,BernevigHughes} associated to the $\mathbb{Z}$ quantized invariant, where the poles correspond to the two Dirac points of the honeycomb lattice. From the same light signals, the quantum Hall conductivity \cite{TKNN} was first introduced through a summation on the wavevectors of the whole Brillouin zone \cite{NathanPeter,DFG,Japan}. 
Our geometrical formulation can also account for interaction effects in the detection of light in a quantitative manner through a variational stochastic approach \cite{PhilippAdolfoKaryn,KaneMelevariational}.
When adding interaction effects on the lattice in the Haldane model, our variational stochastic approach is very efficient to reveal the first-order nature of the quantum phase transition towards the Mott phase \cite{PhilippAdolfoKaryn}.
This approach is also applicable for two-dimensional topological insulators, i.e. the quantum spin Hall (QSH) effect and the Kane-Mele model \cite{KaneMele1,KaneMele2,FuKane}, where circularly polarized light can also detect the $\mathbb{Z}_2$ invariant locally from the Dirac points\cite{KarynLight}. It is important to emphasize that through the effect of a radial magnetic field magnetic monopoles are realized in quantum circuits with applications to the Haldane model \cite{Google,Boulder,Tran}. The quantum metric tensor is also measured associated to phase transitions \cite{Tran}. Within our approach, we have shown that circularly polarized light produces the same effect as a boost of the azimuthal angle building a correspondence between quantum metric and responses to circularly polarized light through the introduction of a geometrical function \cite{KLHReview}. It is also possible to measure the geometrical information through the optical conductivity in quantum materials \cite{BansilFu}.

This is the beginning of this work today. We will then develop the formalism and geometrical responses from the Bloch sphere and the poles down to  the honeycomb Haldane model in the presence of a topological quantum phase transition.
This will be produced from a longitudinal fixed magnetic field $M$ on the sphere corresponding to an alternating potential on the lattice. We will present an algorithm of the method applicable for topological lattice models mapping the surface of the sphere onto the rhomboid Brillouin zone. We present a unification between QAH and QSH effects in materials from coupled planes, leading to a relation with the Ramanujan alternating infinite series and to a $\frac{1}{2}$ topological proximity effect in the thermodynamic limit, related to transport and to the response to circularly polarized light through the Riemann zeta function. 

What are then the precise goals in this article? In recent works, we have shown how the topological invariant can also be re-written in terms of the pseudo-spin responses at the poles related to a protected quantum dynamo effect \cite{dynamo,KLHReview,FractionalArticle} and with applications on topological quantum wires \cite{FrederickLoicKaryn,FrederickLoicOlesiaKaryn,KarynFanMagali}. In this article, in Eq. (\ref{kappaspin}), we introduce the effective magnetic moment of the monopole as a function of $M$ associated to the longitudinal pseudo-spin response. We show how this is related to transport properties. We present the susceptibility for the monopole
 in Eq. (\ref{chitopo}) which remains invariant within the topological phase until the phase transition and therefore can be introduced as a topological marker. To the best of our knowledge, this magnetic susceptibility of the monopole was not mentioned before in the literature; see e.g. Ref. \cite{Alba} where some pseudo-spin responses are introduced related to time of flight in cold atoms. This topological susceptibility shows then an analogy with orbital magnetism where the quantum Hall response is also thought of as a topological orbital susceptibility \cite{SariahKarynFrederic} corresponding to the derivative of the orbital magnetic moment with respect to the applied perpendicular magnetic field. 
The local Berry curvature on the surface of the sphere is also related to the in-plane pseudo-spin response.
 We will then develop the corresponding theory for the Haldane model in two dimensions \cite{Haldane} with the introduction of a Semenoff mass \cite{Semenoff} which corresponds e.g. to the effect of a charge density wave substrate \cite{Eva}. We develop the correspondence between magnetic moment of the monopole and real-space geometrical marker for the Haldane model. The local definition of the topological invariant in terms of Berry gauge potentials and pseudo-spin responses at Dirac points, within our approach, also reveals standard definition in terms of sign of mass at each Dirac point \cite{Haldane,Nathan,Orsay}. Topological properties describing the occurrence of an edge mode are generally robust until the phase transition. Finding observables to describe such a transition towards this trivial phase is yet of great interest related to the relative populations on the two sublattices \cite{Alba}. Indeed, above the transition, particles progressively reside on one site preferably as a result of the proximity effect referring to a trivial insulator which does not present an edge mode and is characterized through a zero index (invariant). It is then relevant to mention here recent progress on measuring locally topological or geometrical properties in momentum space from photo-luminescence related to the Bloch sphere approach \cite{C2N}.
 
 The theory and the developed algorithm will show how the response to circularly polarized light at the Dirac points is efficient to reveal the topological nature of the quantum phase transition i.e. only one Dirac point will resonate with the
 circularly polarized waves leading to a $\frac{1}{2}$ response compared to the topological and trivial phases emphasizing on the $\frac{1}{2}$ topological nature similar to a half Skyrmion at the transition.
Our geometrical approach then will have implications on geometrical properties of coupled-planes materials \cite{DFG2}, alternating the possibility of QSH and QAH effects through the Ramanujan alternating infinite series 
as a physical $\frac{1}{2}$ topological invariant in real space \cite{Review2022}. This will give rise to a topological interface in real space linked to the $\frac{1}{2}$ number that will then relate to our recent works on fractional topological numbers
 that we take time to also remind in the next sentences. Through a model of coupled spins-$\frac{1}{2}$ on the Bloch sphere \cite{FractionalArticle,KLHReview} and two-dimensional materials such as graphene
and bilayer systems \cite{FractionalArticle,KarynSariah1,KarynSariah2,KLHQSHSemimetal}, interacting quantum wires  \cite{FrederickLoicKaryn,FrederickLoicOlesiaKaryn,KarynFanMagali}, the $\frac{1}{2}$ topological
number gives rise to interesting topological phases such as topological semimetals in two dimensions which can also be revealed through circularly polarized light  \cite{FractionalArticle,KarynSariah1,KarynSariah2,KLHQSHSemimetal}, and yet characterizes interacting topological phases in one dimension \cite{FrederickLoicKaryn,FrederickLoicOlesiaKaryn,KarynFanMagali}. 
We underline here recent interesting works in the community related to these efforts \cite{BoFu1,BoFu2,IkedaRayan,Nascimbene}.
This model of two spheres can be realized in quantum circuits \cite{Google} and can be applied as a platform for quantum information through protected Majorana fermions \cite{MajoranaHalf}. The half topological invariant then means that the surface radiating the Berry curvature is halved producing a half Skyrmion and that the geometrical properties are locally resolved at one pole as a $\pi$ Berry phase or $\pi$ winding number \cite{OneHalf}. For the situation of entangled Bloch spheres, this also builds a parallel with the half flux quantum in the sense of superconductivity compared to a metallic phase \cite{OneHalf}.

The organization of the manuscript is as follows.
In Sec. \ref{Monopole}, we develop the geometrical aspects of the monopole in quantum physics associated to the quantum phase transition and we present the evaluation of the spin response and of the
effective magnetic moment as a function of the magnetic field $M$. We also present the relation with the (mean) quantized transverse Hall current from a Newtonian approach. 
In Sec. \ref{maplattice}, we develop the theory for the honeycomb Haldane model in the presence of a Semenoff mass producing the quantum phase transition. We present a precise map from the rhomboid Brillouin zone onto the sphere and derive analytically and numerically the local responses associated to local invariants in momentum (reciprocal) space. The spin-1/2 on the Bloch sphere will correspond to the pseudo spin-1/2 measuring the population imbalance between the two sublattices of the honeycomb lattice. We introduce a real space analysis related to the effective magnetic moment of the monopole. We elaborate on the local responses from circularly polarized light associated to the quantum
phase transition. In Sec. \ref{coupledplanes}, we develop the theory of coupled-planes related to the $\frac{1}{2}$ topological invariant building an analogy between a topological interface and the Ramanujan infinite alternating series.
We analyse transport properties and the responses to circularly light within this correspondence. The coupled-planes model includes the alternating presence of QAH and QSH effects. In Sec. \ref{summary}, we summarize our findings.
Appendices are devoted to additional mathematical proofs. In Appendix \ref{AppendixA}, we elaborate on the relation between the geometrical approach and transport, quantum Hall response with a map onto a cylinder.
In Appendix \ref{AppendixB}, we present a geometrical justification of the $\frac{1}{2}$ topological invariant for the coupled planes when reaching the infinity limit.

\section{Bloch Sphere as a Magnetic Monopole, Quantum Hall Response and Topological Magnetism}
\label{Monopole}

The magnetic monopole in quantum physics is then formed through the application of a radial magnetic field associated to the Bloch sphere of a spin-1/2, resulting in a hedgehog topological sphere \cite{KLHReview}. 
We remind here that such an Hamiltonian is realized in quantum circuits \cite{Google,Boulder,Tran}. 

In this Section, we study the physical responses as a function of the parameter $M$ corresponding to the fixed additional magnetic field along $z$ direction which will drive the quantum phase transition associated to the Poincar\' e-Hopf theorem. We evaluate the local spin magnetization responses at and in the vicinity of the poles on the Bloch sphere as a function of $M$. 
We present a simple proof in a dressed polar angle representation showing how the global topological number is generally measurable locally through the Berry gauge potential \cite{KLHReview,FractionalArticle} 
which can then be re-written in terms of the spin magnetization along $z$ direction through Eherenfest theorem. We will then introduce the mean (medium) of the spin magnetization on the surface area and address the relation with the transverse quantum Hall current or transverse pumped charge in a Newtonian (gravitational) correspondence when driving a charge from north to south pole through a longitudinal electric field. This will give rise to the effective topological magnetic moment $\kappa(M)$ in Eq. (\ref{kappaspin}), associated to the loops of transverse currents formed along the equatorial plane on the whole surface. When $M=0$, as a result of the radial magnetic field, magnetic moments of the two hemispheres compensate each other and $\kappa(M=0)=0$. This analysis will also be related to the cylinder geometry \cite{KLHReview}.

It should be emphasized that for classical magnetic monopoles produced e.g. by the same radial magnetic field, the magnetic field $M$ corresponds to an additional source to the Gauss theorem that produces an additional effective zero flux on a spherical surface i.e. the in-coming magnetic flux at south pole is compensated by the out-going magnetic flux at north pole. In that case, the $M$ term will not mediate a classical phase transition.

\subsection{Topological Bloch Sphere and Geometrical Approach of the Quantum Phase Transition through a Local Spin Marker}

We begin with the quantum Hamiltonian for the spin-1/2 
\begin{equation}
H = - (B{\bf e}_r +M{\bf e}_z)\cdot \boldsymbol{\sigma}.
\end{equation}
Here, the vector $\boldsymbol{\sigma}=(\sigma_x,\sigma_y,\sigma_z)$ is associated to Pauli matrices.
The total magnetic field corresponds to the addition of a radial magnetic field ${\bf B}$ and of a magnetic field $M$ polarized along $z$ direction. Then, ${\bf e}_r$ and ${\bf e}_z$ are two unit vectors
associated to these directions. The Hamiltonian can be equivalently written in spherical coordinates as
\begin{equation}
H = - {\bf B}^*\cdot \boldsymbol{\sigma}
\end{equation}
with the magnetic field 
\begin{equation}
{\bf B}^* = B\left(\sin\theta \cos \varphi, \sin\theta \sin \varphi, \cos \theta+\frac{M}{B}\right).
\end{equation}

\begin{figure}[t]
\includegraphics[width=1\textwidth]{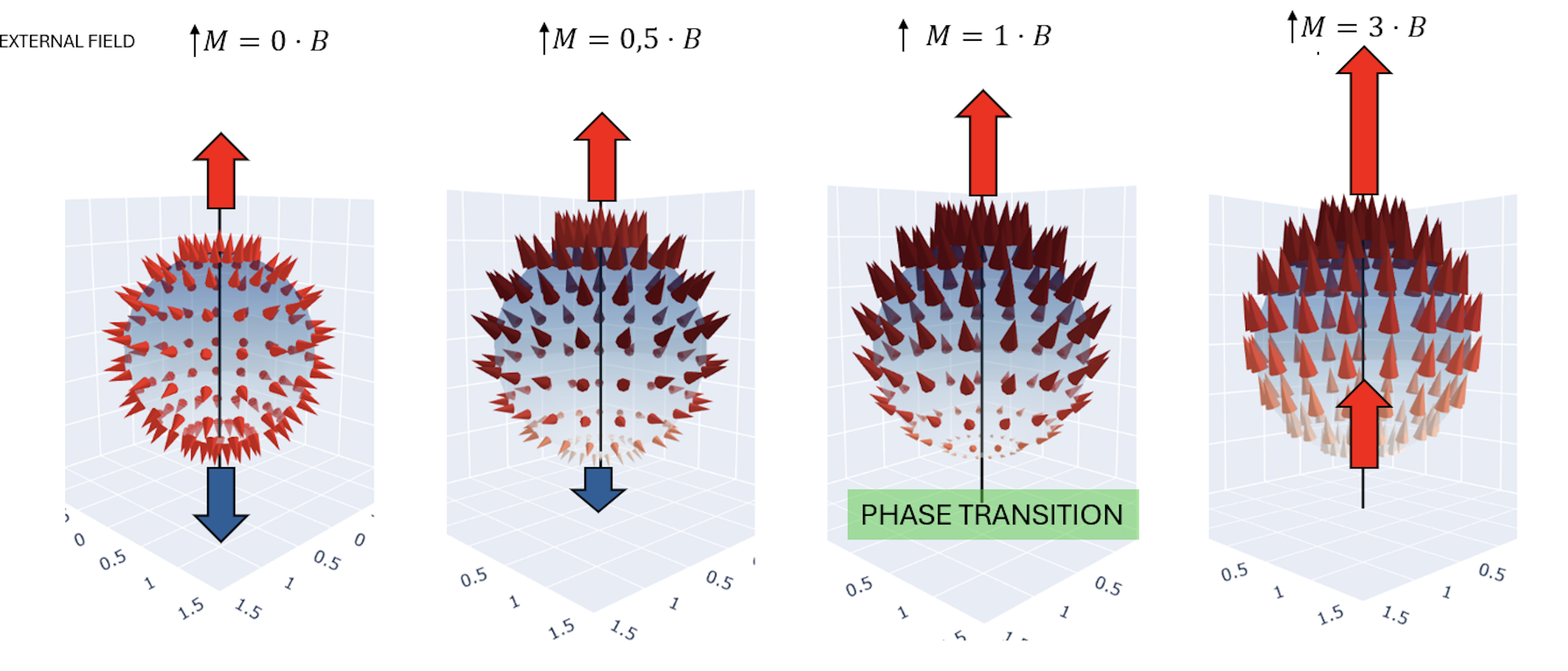} 
\vskip -0.3cm
\caption{Local variation of the ${\bf n}=\frac{{\bf B}^*}{B^*}$ vector associated to the total magnetic field as a function of $M$ on the unit sphere associated to the angles $\theta$ and $\varphi$. Phase transition refers here to the situation of the topological quantum phase transition occurring when $M=B$ i.e. it will be associated to the jump of the topological invariant from one to zero. Associated to this transition, the $n_z$-component of this unit vector flips its direction at south pole.}
\label{spinresponse}
\end{figure}

Here, $\theta\in [0;\pi]$ and $\varphi\in [0;2\pi]$ are the polar and azimuthal angles respectively in spherical coordinates.
In Fig. \ref{spinresponse}, we present the local variation of the unit vector ${\bf n}=\frac{{\bf B}^*}{B^*}$ as a function of $M$ on the unit sphere associated to the angles $\theta$ and $\varphi$.
The energy eigenvalues are
\begin{equation}
E_{\pm} = \pm \sqrt{(B\cos \theta +M)^2 + B^2 \sin^2 \theta} = \pm B^* = \pm |{\bf B}^*|.
\end{equation}
The eigenstate with (lowest) energy $E_-=-B^*$ takes the form
\begin{equation}
|\psi\rangle = |\alpha| e^{-i \frac{\varphi}{2}}|+\rangle + |\beta| e^{i \frac{\varphi}{2}}|-\rangle
\end{equation}
where
\begin{eqnarray}
|\alpha|^2 &=& (E_- - (B\cos \theta+M))/(2 E_-) \nonumber \\
|\beta|^2 &=& (E_- + (B\cos \theta+M))/(2 E_-).
\end{eqnarray}
The Hilbert space is formed with the 2D eigenstates $|+\rangle$, $|-\rangle$ associated to the $z$ direction (or Pauli matrix $\sigma_z$) with spin eigenvalues $+1$ and $-1$ respectively.
The eigenstates are formulated within a particular gauge choice for the phase related to the azimuthal angle. The theory presented below will be gauge invariant.

\begin{figure}[t]
\includegraphics[width=0.9\textwidth]{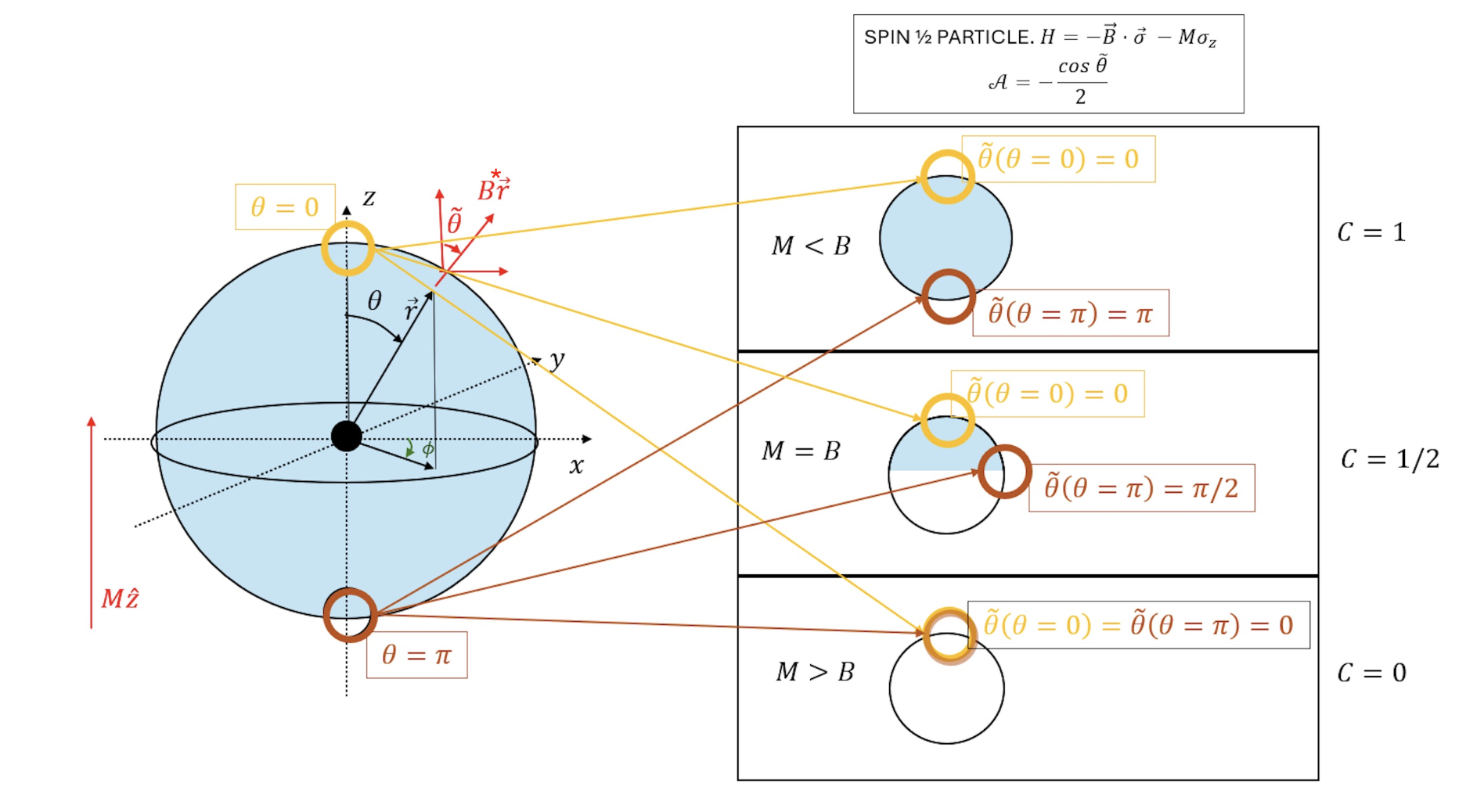} 
\vskip -0.3cm
\caption{(Left) Representation of the total magnetic field ${\bf B}^*$ on the unit sphere described through the angles $\theta$ and $\varphi$. (Right) Representation of the effective topological properties through the (dressed) angle $\tilde{\theta}(\theta)$. The direction of the total magnetic field ${\bf B}^*$ is then radial for all the values of $M$ associated to the unit vector $\bf{n}^*$ in Eq. (\ref{effectivedirection}). We introduce $\mathcal{A}=A_{\varphi}$ in Eq. (\ref{Aphi}).
For all the phases the north pole is identified with $\theta=0$ i.e. $\tilde{\theta}=0$. Within the topological phase $B<M$, the south pole corresponds to $\tilde{\theta}(\theta=\pi)=\pi$. For $M<B$ topological properties are equivalent to those of a Skyrmion. For $M>B$, the north and south poles are identical and the equivalent geometry does not encircle the topological charge at the origin. For $M=B$, geometrical properties are equivalent to a half sphere corresponding to a half monopole or a half Skyrmion.}
\label{anglecorrespondence}
\end{figure}

In quantum physics the Berry curvature plays the role of the magnetic field and the equivalent to the vector potential is the local Berry gauge potential \cite{Berry}. The topological invariant is measured from the integral of the Berry curvature on the surface. It equivalently measures a topological charge in the core of the sphere. It can be equivalently introduced in spherical coordinates or through a cartesian metric from the simplification of the $\sin \theta$ function in the differential area and the $\frac{1}{\sin\theta}$ factor stemming in the definition of the Berry gauge potential component along the equatorial direction in spherical coordinates \cite{Google}. In this way, we introduce the Berry gauge potential as
\begin{equation}
A_{\varphi} = -i\langle \psi| \partial_{\varphi} |\psi\rangle = -\frac{1}{2}(|\alpha|^2-|\beta|^2) = \frac{1}{2 E_-}(B\cos\theta+M).
\end{equation}
We also have that $A_{\theta} = -i \langle \psi| \partial_{\theta} |\psi\rangle = 0$. The Berry curvature then takes the form $F_{\theta \varphi} = \partial_{\theta} A_{\varphi} -
\partial_{\varphi} A_{\theta} = \partial_{\theta} A_{\varphi}=\frac{\partial A_{\varphi}}{\partial \theta}$. We can then introduce the topological invariant as 
\begin{equation}
C = \frac{1}{2\pi}\int_0^{2\pi} d\varphi\int_0^{\pi} F_{\theta \varphi} d\theta = \int_0^{\pi} \frac{\partial A_{\varphi}(\theta,M)}{\partial \theta} d\theta.
\end{equation}
The differential associated to $A_{\varphi}$ depends on the parameter $M$. Therefore to simplify this formula, it is useful to introduce the {\it dressed angle}
\begin{equation}
\tan \tilde{\theta} = \frac{\sin \theta}{B \cos \theta + M}.
\end{equation}
In principle, the general correspondence should be thought of as $B^*\cos \tilde{\theta}=B\cos \theta + M$ and we introduce $B^*\sin\tilde{\theta}=B\sin\theta$. On the sphere described through the dressed angle $\tilde{\theta}$ and
$\varphi$, the magnetic field is radial ${\bf B}^*=B^*{\bf e}_r$. The Berry gauge potential and the Berry curvature will not depend specifically on the precise form of $B^*$. Therefore, we equivalently introduce the unit radial ${\bf n}^*$ vector for this geometrical correspondence in Fig. \ref{anglecorrespondence}(Right)
\begin{equation}
\label{effectivedirection}
{\bf n}^* = (\sin\tilde{\theta}\cos\varphi,\sin\tilde{\theta}\sin\varphi,\cos\tilde{\theta}).
\end{equation}

In this way, the topological invariant reads
\begin{eqnarray}
\label{invariant}
C &=& \int_{\tilde{\theta}(\theta=0)}^{\tilde{\theta}(\theta=\pi)} \frac{\partial A_{\varphi}(\tilde{\theta})}{\partial \tilde{\theta}}\frac{\partial \tilde{\theta}}{\partial \theta} d{\theta} = 
\int_{\tilde{\theta}(\theta=0)}^{\tilde{\theta}(\theta=\pi)} \frac{\partial A_{\varphi}(\tilde{\theta})}{\partial \tilde{\theta}} d\tilde{\theta}. \nonumber \\
&=& A_{\varphi}(\tilde{\theta}(\theta=\pi)) - A_{\varphi}(\tilde{\theta}(\theta=0)),
\end{eqnarray}
where
\begin{equation}
\label{Aphi}
A_{\varphi}(\tilde{\theta}) = - \frac{1}{2}\cos\tilde{\theta}.
\end{equation}
The last equation is equivalent to $|\alpha|^2=\cos^2\frac{\tilde{\theta}}{2}$ and $|\beta|^2=\sin^2\frac{\tilde{\theta}}{2}$. The topological invariant is measured for any fixed value of $M$. This local formulation of the topological invariant from the poles is gauge invariant and it agrees with general geometrical thoughts allowing for a correspondence on the cylinder \cite{KLHReview,FractionalArticle} (see Appendix \ref{AppendixA}). All the dependence on the parameter $M$ is then hidden into the definitions of the angles $\tilde{\theta}(\theta=0)$ and $\tilde{\theta}(\theta=\pi)$. We emphasize that the eigenstates are introduced within the same coherent gauge on the whole surface.

As shown in Fig. \ref{anglecorrespondence},  for all values of $M$, 
the north pole associated to $\theta=0$ is also equivalent to $\tilde{\theta}=0$ i.e. $\tilde{\theta}(\theta=0)=0$. On the other hand, we do observe a quantum phase transition when $M=B$ from the angle $\tilde{\theta}$: as long as $M<B$, then $\tilde{\theta}=\pi$ also corresponds to south pole i.e. to $\theta=\pi$ such that $\tilde{\theta}(\theta=\pi)=\pi$, when $M=B$ the topological properties become equivalent to a half of a sphere $\tilde{\theta}(\theta=\pi)=\frac{\pi}{2}$ and for $M>B$ we also have the important equality 
$\tilde{\theta}(\theta=\pi)=\tilde{\theta}(\theta=0)$ implying $C=0$. The topological number then jumps from one when $M<B$ to zero when $M>B$. Within this formulation, at the quantum phase transition we also identify
\begin{equation}
C_{1/2} =  - A_{\varphi}(\tilde{\theta}(\theta=0)) = +\frac{1}{2},
\end{equation}
which can be interpreted as a half Skyrmion. It is interesting to mention that $A_{\varphi}(\tilde{\theta}(\theta=0))=-\frac{1}{2}$ corresponds to a topological characterization of a $+\pi$ Berry phase encircling the north pole \cite{FractionalArticle,OneHalf}.
The classical analogue of the local Berry gauge potential i.e. the classical vector potential reads ${\mathcal A}=A_{\varphi}(M)=-B\frac{\cos\theta}{2}-\frac{M}{4}\cos(2\theta)$ when setting $r=1$. 
This generalizes the formula introduced in Eq. (10) in Ref. \cite{KLHReview} for $M=0$.  We verify that {\it classically} the induced $M$-component $-\frac{M}{4}\cos(2\theta)$ remains identical at the two poles 
justifying why the Gauss theorem only measures the presence of the radial magnetic field for this situation.

From Ehrenfest's theorem $\langle \psi|\sigma_z |\psi \rangle = \langle \sigma_z\rangle = \cos\tilde{\theta}=-2A_{\varphi}$. Then, the topological number can be reformulated through the spin polarizations at the north pole and at the dressed angle associated to 
$\theta=\pi$:
\begin{equation}
\label{topo}
C = \frac{1}{2}(\langle \sigma_z(\tilde{\theta}(\theta=0))\rangle - \langle \sigma_z(\tilde{\theta}(\theta=\pi))\rangle).
\end{equation}
This definition agrees with previous results on one sphere when $M=0$ \cite{dynamo}. 
This formula was adapted for the analysis of topological quantum wires with applications in real space associated to correlation functions \cite{FrederickLoicKaryn,FrederickLoicOlesiaKaryn,KarynFanMagali}.
The profile of the magnetic structure at the poles in Fig. \ref{spinresponse} clearly reveals the quantum phase transition.
At the topological quantum phase transition, we have the identity
\begin{equation}
\label{onehalf}
C_{1/2} = \frac{1}{2} = \frac{1}{2} \langle \sigma_z(\theta=0)\rangle.
\end{equation}
For two spheres, it is possible to reach fractional $\frac{1}{2}$ numbers on a line or in a phase of the parameters space associated to the interaction between spins \cite{KLHReview}.
The local formalism between Berry gauge potential, pseudo-spin physics is also able to measure the existence of a Bell state or Einstein-Podolsky-Rosen pair at one pole \cite{OneHalf}.
For $M>B$, the spin magnetizations associated to $\theta=0$ and $\theta=\pi$ are the same such that $C=0$ i.e. $\tilde{\theta}(\theta=0)=\tilde{\theta}(\theta=\pi)$.

In superconducting quantum circuits, the topological invariant is measured from the Berry curvature when driving from north to south pole \cite{Google,Boulder}. The topological quantum phase transition was also revealed in this way. We propose
a local alternative representation of the topological invariant and of the topological phase transition which is also related to the physical response of the spin at the poles.

Since the longitudinal and in-plane pseudo-spin responses are related through a partial derivative with respect to $\tilde{\theta}$, $|\sigma_x(\tilde{\theta})|$ is then directly related to the local Berry curvature 
$F_{\tilde{\theta}\varphi}=
\partial_{\tilde{\theta}}A_{\varphi}$ such that
we also have
\begin{equation}
C = \frac{1}{2}\int_{\tilde{\theta}(\theta=0)}^{\tilde{\theta}(\theta=\pi)} |\sigma_x(\tilde{\theta})| d\tilde{\theta}.
\end{equation}

\subsection{Map of Spin Response on the Area and Effective Topological Magnetic Moment, Quantized Transverse Hall current from a Newtonian Approach}
\label{sphereresponse}

In Appendix \ref{AppendixA}, we generalize the proof of Refs. \cite{KLHReview,FractionalArticle} from the Parseval-Plancherel theorem showing the induction of a quantized transverse pumped current in the {\it dressed polar angle} representation
on the sphere
\begin{equation}
J_{\perp}(\tilde{\theta}) = \frac{Q_{\perp}}{T} = \frac{1}{2\pi}\oint J_{\perp}(\tilde{\theta})d\varphi.
\end{equation}
The charge $e$ is the charge of an electron. The pumped charge {\it at an angle $\tilde{\theta}$} reads
\begin{equation}
Q_{\perp}(\tilde{\theta}) = \sin^2\frac{\tilde{\theta}}{2}.
\end{equation}
In this formulation, in the presence of the radial magnetic field acting on the spin of an electron and in the presence of the magnetic field source $M$, we measure the transverse pumped charge associated to an electric field applied along the polar angle such that $\hbar\dot{\tilde{\theta}}=e{\mathcal E}$. At time $t=0$, a charge is present at north pole. The charge goes down similar to the apple and the gravitational force is a result of the electric field directed along the polar angle direction i.e. it corresponds to a Coulomb force. The electron moves on the surface of this ball associated to the Bloch sphere. The final time $T$, corresponding to the time of the measure, is fixed to be identical for any $M$ such that the angle $\tilde{\theta}$ reaches $\pi$ i.e. $T=\frac{e{\mathcal E}}{\hbar \pi} = \frac{2e{\mathcal E}}{h}$. Within the topological phase, the transverse pumped charge at south pole precisely measures the topological invariant itself $C=1$. See Appendix \ref{AppendixA}. When we reach the transition maintaining the same value of the (final) time $T$ for the measure, associated to Fig. \ref{anglecorrespondence}, this corresponds to maintain the Berry gauge potential fixed to $A_{\varphi}(\tilde{\theta}=\pi)=A_{\varphi}(\tilde{\theta}=\frac{\pi}{2})=0$ for dressed angles $\tilde{\theta}\in [\frac{\pi}{2},\pi]$ such that there is no Berry curvature effectively present in the south hemisphere. In that case, the topological invariant becomes halved i.e. equal to $C_{1/2}$ and the transverse pumped
charge is also halved. Above the transition, the charge remains at the north pole i.e. $A_{\varphi}(\tilde{\theta}=\pi)=A_{\varphi}(\tilde{\theta}=0)$ and there is no transverse pumped current. 
In Appendix  \ref{AppendixA}, we also show the relation with the cylinder geometry. 

In this article, then we introduce the {\it mean} value of the transverse pumped charge on the original sphere associated to the polar angle $\theta\in [0;\pi]$
\begin{equation}
\bar{Q}_{\perp} = \frac{e}{4\pi}\int_0^{2\pi} d\varphi \int_0^{\pi} Q_{\perp}(\tilde{\theta}) \sin \theta d\theta=\frac{e}{4\pi}\int_0^{2\pi} d\varphi \int_0^{\pi} \sin^2\frac{\tilde{\theta}}{2} \sin \theta d\theta.
\end{equation}
The justification is that when $M=0$, we find that this response also acquires a topological origin
\begin{equation}
\bar{Q}_{\perp} = \frac{e C(M=0)}{2}.
\end{equation}
When $M\neq 0$, we introduce the formula
\begin{equation}
\bar{Q}_{\perp} = \frac{e C(M=0)}{2} - \frac{e\kappa(M)}{2},
\end{equation}
with
\begin{equation}
\label{kappaspin}
\kappa(M) = \frac{1}{4\pi}\int_0^{2\pi} d\varphi \int_0^{\pi} \langle \sigma_z(\tilde{\theta})\rangle \sin \theta d\theta = \frac{1}{2}\int_0^{\pi}  \langle \sigma_z(\tilde{\theta})\rangle \sin \theta d\theta = \frac{I}{2}.
\end{equation}
Here, $\kappa$ is the mean value of the spin response along $z$ direction or the effective spin response as a function of $M$.
When $M=0$, we emphasize here that from symmetry between the two hemispheres $\kappa=0$. In response to the radial magnetic field in Fig. \ref{spinresponse} the two hemispheres produce an effective moment equal in amplitude but opposite in directions.
When $M\rightarrow +\infty$, $\langle \sigma_z(\tilde{\theta})\rangle\rightarrow 1$ such that the mean value of the transverse pumped charge on the whole area is zero. 
In that limit, $\kappa$ is a measure of the unit area.
We are precisely questioning the (topological) properties of $\kappa$ as a function of $M$ when inserting the equations
\begin{equation}
\label{sigmaz}
\langle \psi | \sigma_z | \psi\rangle = \cos\tilde{\theta} = \frac{B\cos\theta+M}{\sqrt{B^2\sin^2 \theta + (B\cos\theta+M)^2}} = \frac{\partial B^*}{\partial M}=-\frac{\partial E_-}{\partial M}.
\end{equation}

\begin{figure}[t]
\includegraphics[width=0.7\textwidth]{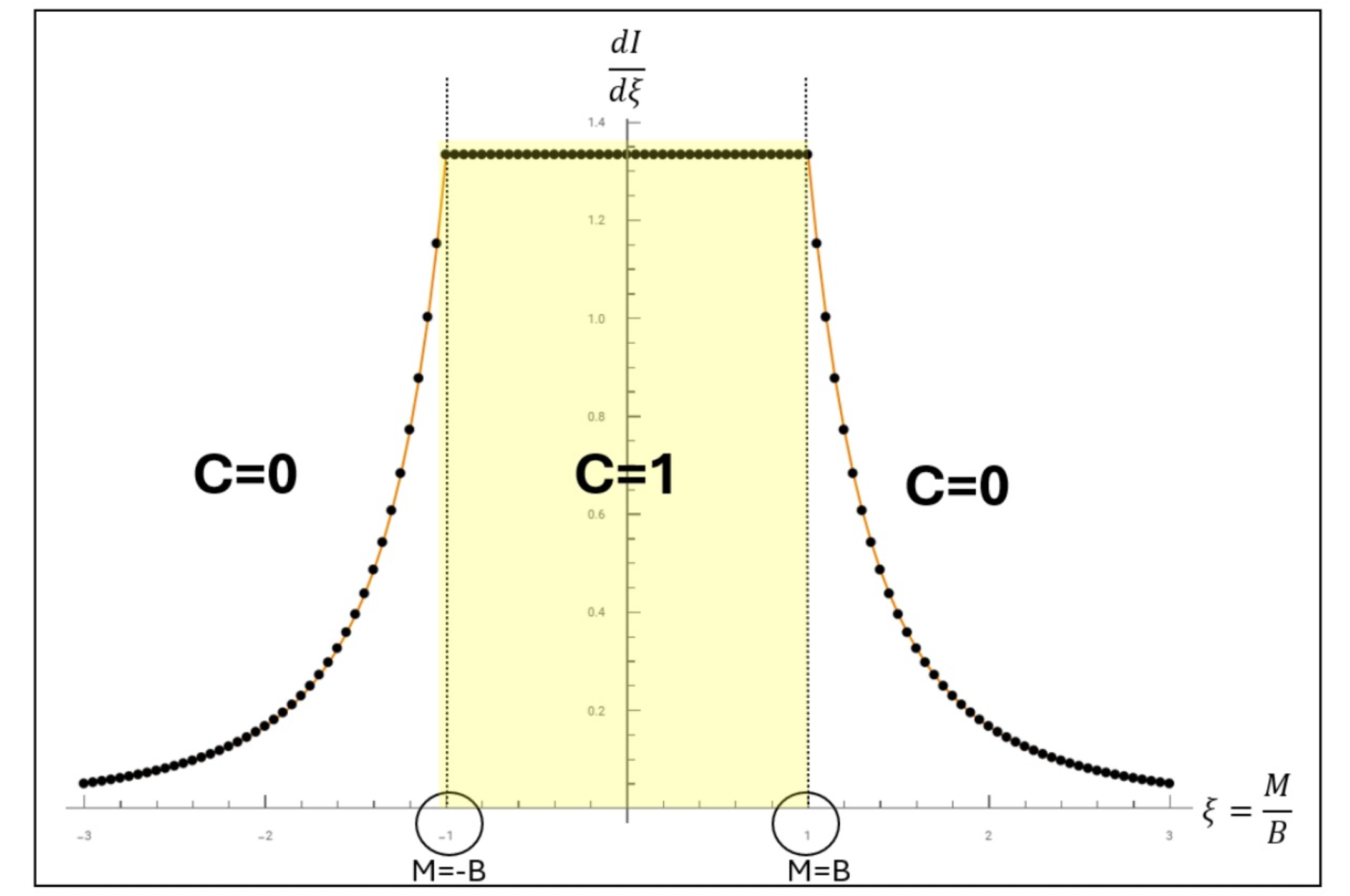} 
\vskip -0.3cm
\caption{Numerical evaluation of $\frac{dI}{d\xi}=2\chi$ as a function of $\xi$ that reproduces the analytical results, in particular the quantized plateau for the susceptibility within the topological phase.}
\label{pseudospinsphere}
\end{figure}

To evaluate $\kappa(M)$, we will first write down a correspondence between spin response and magnetic field. 
We introduce the definition
\begin{equation}
B^* = B\sqrt{1+x}
\end{equation}
with
\begin{equation}
x=\left(2\frac{M}{B}\cos\theta+\frac{M^2}{B^2}\right).
\end{equation}
At fixed $M$, $2M d(\cos \theta) = B dx$. 
In this way,
\begin{equation}
\kappa(M) = -\frac{B}{2}\frac{\partial}{\partial M} \left( \frac{1}{2M}\int_{x_{min}=x(\theta=0)}^{x_{max}=x(\theta=\pi)} B^* dx\right).
\end{equation}
This is equivalent to
\begin{equation}
\label{equation}
\kappa(M) = -\frac{B^2}{2}\frac{\partial}{\partial M} \left(\frac{1}{2M} \int_{x_{min}=x(\theta=0)}^{x_{max}=x(\theta=\pi)} \sqrt{1+x}\ dx\right).
\end{equation}
Then, we can relate $x_{min}=2\frac{M}{B}+\frac{M^2}{B^2}$ and $x_{max}=-2\frac{M}{B}+\frac{M^2}{B^2}$ revealing the properties of the magnetic field $\bf{B}^*$ at the poles in Fig. \ref{spinresponse}:
\begin{equation}
\sqrt{1+x_{max}} = \frac{B^*(\theta=\pi)}{B}= \left|1-\frac{M}{B}\right|
\end{equation}
\begin{equation}
\sqrt{1+x_{min}} = \frac{B^*(\theta=0)}{B}= \left(1+\frac{M}{B}\right).
\end{equation}
This leads to
\begin{equation}
\label{formulaspin}
\kappa(M) = -\frac{B^2}{2}\frac{\partial}{\partial M} \left(\frac{1}{3M} \left|1-\frac{M}{B}\right|^3 - \frac{1}{3M}\left(1+\frac{M}{B}\right)^3 \right).
\end{equation}
For $M<B$, the only terms which are relevant are the quadratic terms in the parenthesis which then turn into a linear term when applying the partial derivative with respect to $M$.
For $M<B$, we find
\begin{equation}
\kappa(M) = \frac{M}{B}\frac{2}{3}.
\end{equation}
When $M<B$, this results in a quantized plateau for the susceptibility response. The linear response regime then remains applicable until the transition point. We verify this result with a numerical integration of $\frac{dI}{d\xi}$
with $I=2\kappa$ and $\xi=\frac{M}{B}$ in Fig. \ref{pseudospinsphere}. A nice result of this article is then to introduce a {\it topological magnetic susceptibility response} for the hedgehog sphere (Skyrmion qubit) within the topological phase
\begin{equation}
\chi = B\frac{\partial \kappa}{\partial M} = \frac{2}{3}.
\end{equation}
The susceptibility has indeed a topological origin when $M<B$. To show this we can re-phrase the result when $M<B$ as 
\begin{equation}
\kappa = -\frac{1}{6B} \frac{\partial}{\partial M}\left(\left(1-\frac{M}{B}\right)M\right) + \frac{1}{6B} \frac{\partial}{\partial M}\left(\left(1+\frac{M}{B}\right)M\right).
\end{equation}
This is then equivalent to
\begin{equation}
\label{chitopo}
\chi= \chi_{topo} = B\frac{\partial \kappa}{\partial M} = \frac{2}{3}C(M<B) = -2B\frac{\partial \bar{Q}_{\perp}}{\partial M},
\end{equation}
where the topological invariant is the invariant in the topological phase which can be written as
\begin{equation}
C(M<B) = \frac{1}{2}(\langle \sigma_z(0)\rangle - \langle \sigma_z(\pi)\rangle)=1
\end{equation}
from Eq. (\ref{topo}) and we insert the definition in Eq. (\ref{sigmaz}) for the spin responses. 
Within the topological phase $M=B^-$ the topological character of $\kappa$ can be summarized as
\begin{equation}
\kappa = \kappa_{topo} = -\frac{1}{6B}\left(B^*(\theta=\pi)-B^*(\theta=0)\right) + \frac{M}{3B}C(M<B).
\end{equation}
Since
\begin{equation}
\frac{1}{2}\left(B^*(\theta=0)-B^*(\theta=\pi)\right) = MC(M<B)
\end{equation}
then, this also leads to
\begin{equation}
\kappa = \kappa_{topo} = \frac{2M}{3B}C(M<B).
\end{equation}
This formula is valid until $M=B^-$. 
For $M>B$, the form of $\kappa$ in Eq. (\ref{formulaspin}) is
\begin{equation}
\kappa(M) = 1-\frac{1}{3}\frac{B^2}{M^2},
\end{equation}
which leads to a power-law susceptibility response in $M^{-3}$. For $M=B$, the mean value of the magnetic moment and the susceptibility are continuous. 

Then, we present another view of this proof related to the properties of the topological invariant within the topological phase. We develop the $\sqrt{1+x}$ function in Eq. (\ref{equation}) as an infinite series and we develop the response order by order in $x=2\frac{M}{B}\cos\theta+\frac{M^2}{B^2}$. To order $\frac{x}{2}$, this leads to
\begin{equation}
\kappa_{\frac{x}{2}} = \frac{M}{B}C(M=0)
\end{equation}
with the topological invariant 
\begin{equation}
C(M=0) = \frac{1}{2}\int_0^{\pi} \sin\theta d\theta=\frac{1}{2}\int_0^{\pi} \sin\tilde{\theta} d\tilde{\theta}=+1.
\end{equation}
The second equality reveals $C(M=0)=C(M<B)$ and imply that we study the topological phase. The linear term in $M$ in $\kappa$ coming from $-\frac{x^2}{8}$ in the series associated to $\sqrt{1+x}$ then reads
\begin{equation}
\kappa_{-\frac{x^2}{8}} = -\frac{1}{3}\frac{M}{B}C(M=0).
\end{equation}
Therefore, 
\begin{equation}
\kappa_{\frac{x}{2}} + \kappa_{-\frac{x^2}{8}} = \frac{2}{3} \frac{M}{B} C(M=0).
\end{equation}
Then, we verify that this result is in fact robust when including higher order terms in the series. This requires a systematic approach that the numerical method also reproduces. Therefore, we verify in this way that
\begin{equation}
\kappa(M<B)=\kappa_{\frac{x}{2}} + \kappa_{-\frac{x^2}{8}} = \frac{2}{3} \frac{M}{B} C(M<B) = \frac{2}{3}\frac{M}{B}.
\end{equation}

These results can be directly tested in quantum circuits \cite{Google, Boulder, Tran}.

\section{Correspondences on a 2D Topological Lattice Model}
\label{maplattice}

We discuss below an application for the topological Haldane model on the honeycomb lattice \cite{Haldane} that can be realized with circularly polarized light \cite{Cavalleri} and may be engineered in transition metal dichalcogenide (TMD) materials \cite{Mak}.

On the honeycomb 2D lattice, the analogous quantity to $\kappa(M)$ reads
\begin{equation}
\langle \sigma_z({\bf R}_i) \rangle = \frac{1}{N}\sum_{\bf k} \langle \sigma_z({\bf k})\rangle.
\end{equation}
Here, $N=N_A=N_B$ corresponds to the number of $A$ or $B$ inequivalent sites and $\langle \sigma_z\rangle$ measures the difference of occupancies on $A$ and $B$ sublattices. 
In Sec. \ref{realspace}, we will study the properties of this local marker $\langle \sigma_z({\bf R}_i) \rangle$ on the lattice as a function of $M$. 

\subsection{Map from Brillouin Zone of Honeycomb Lattice onto the Sphere and Local Topological Responses for the Haldane Model}
\label{Map}

The monopole formalism \cite{KLHReview} can be developed to analyse properties of lattice models such as the Haldane model on the honeycomb lattice \cite{Haldane}. 

We develop the theory from the Brillouin zone onto the sphere and introduce a numerical algorithm for the evaluation of physical properties in momentum (reciprocal) space and
in real space. The Hamiltonian can be written as a $2\times 2$ matrix in the spinor representation associated to sublattices $A$ and $B$, $\{|{\bf k}, A\rangle, |{\bf k}, B\rangle\}$, such that $H=\sum_{\bf k} H({\bf k})$ where $H({\bf k})=-{\bf d}\cdot \boldsymbol{\sigma}$ and
${\bf d}=(d_x,d_y,d_z+M)$ with
\begin{eqnarray}
\label{dvector}
d_x &=& t\left(1+2\cos\left(\frac{3a}{2}k_x\right)\cos \left(\frac{\sqrt{3}a}{2}k_y\right)\right) \nonumber \\
d_y &=& 2t \sin \left(\frac{3a}{2}k_x\right) \cos\left(\frac{\sqrt{3}a}{2}k_y\right) \nonumber \\
d_z &=& 2t_2\left(\sin \left(\sqrt{3}a k_y\right) - 2\cos\left(\frac{3a}{2}k_x\right)\sin\left(\frac{\sqrt{3}a}{2} k_y\right)\right).
\end{eqnarray}
The term $t$ corresponds to the hopping of electrons on nearest-neighboring sites and $t_2$ is the Haldane hopping term between second nearest neighbors with a complex phase fixed to $\frac{\pi}{2}$ \cite{Haldane}. 
Here, $a$ is the lattice spacing. The two Dirac points have the same component $K_x=\frac{2\pi}{3a}$ such that $\cos(\frac{3a}{2}K_x)=-1$. We also have $K_y=\frac{2\pi}{a3\sqrt{3}}$ and $K'_y=-\frac{2\pi}{a3\sqrt{3}}$ such that
close to the Dirac points we have the identifications \cite{Haldane}
\begin{equation}
\label{dz}
d_z({\bf K}^{\zeta}) = \pm 3\sqrt{3}t_2 = \zeta 3\sqrt{3}t_2.
\end{equation}
The symbol $\zeta$ will be fixed such that $\zeta=+1$ at the $K$ Dirac point and $\zeta=-1$ at the $K'$ Dirac point which can be viewed as the mass inversion effect at the two Dirac points. 
The parameter $M$, favoring the occupancy of a particle (electron) on one sublattice, is called a Semenoff mass term related to the Dirac equation \cite{Semenoff}.

\begin{figure}[t]
\includegraphics[width=1.05\textwidth]{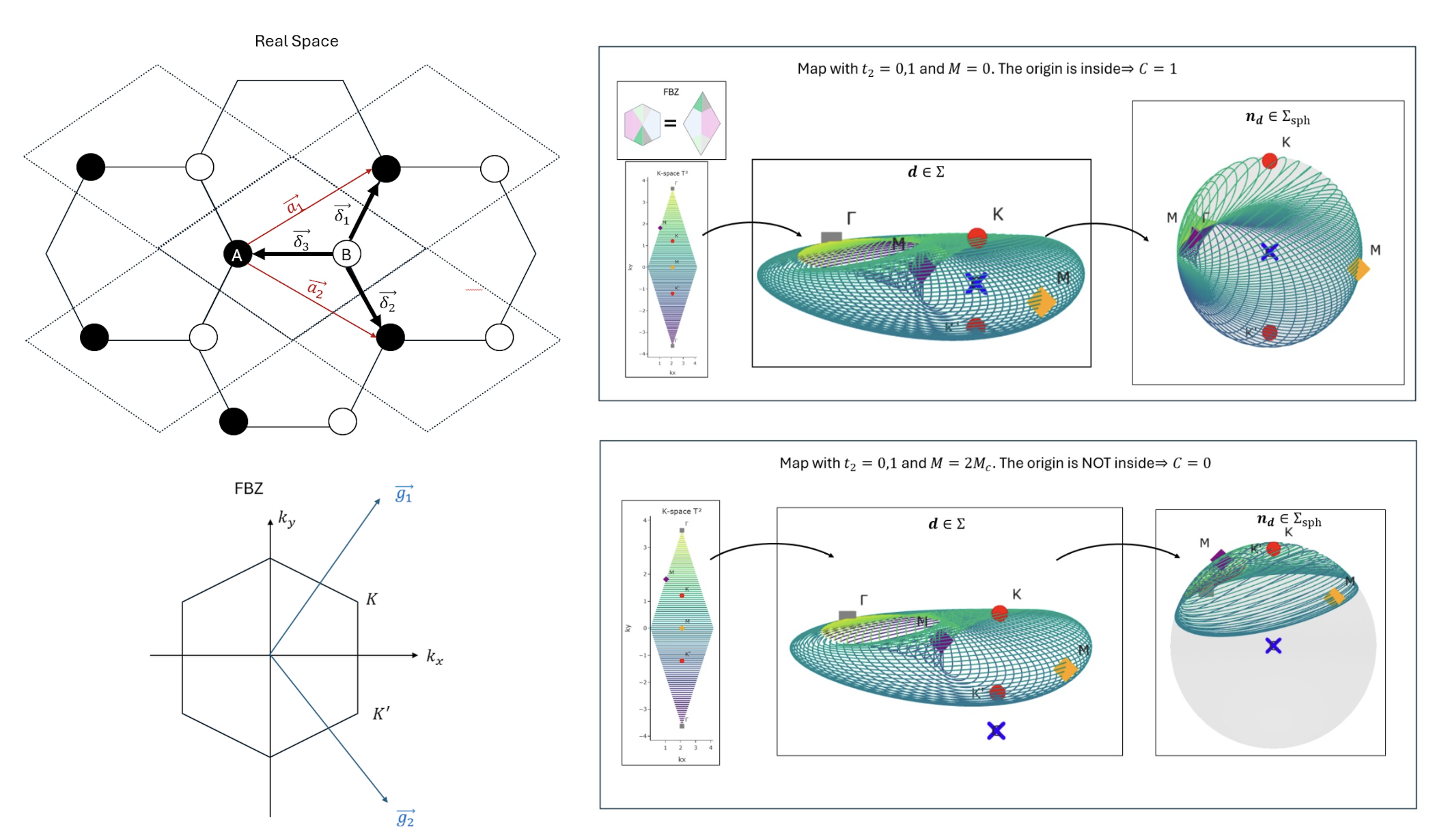} 
\includegraphics[width=0.9\textwidth]{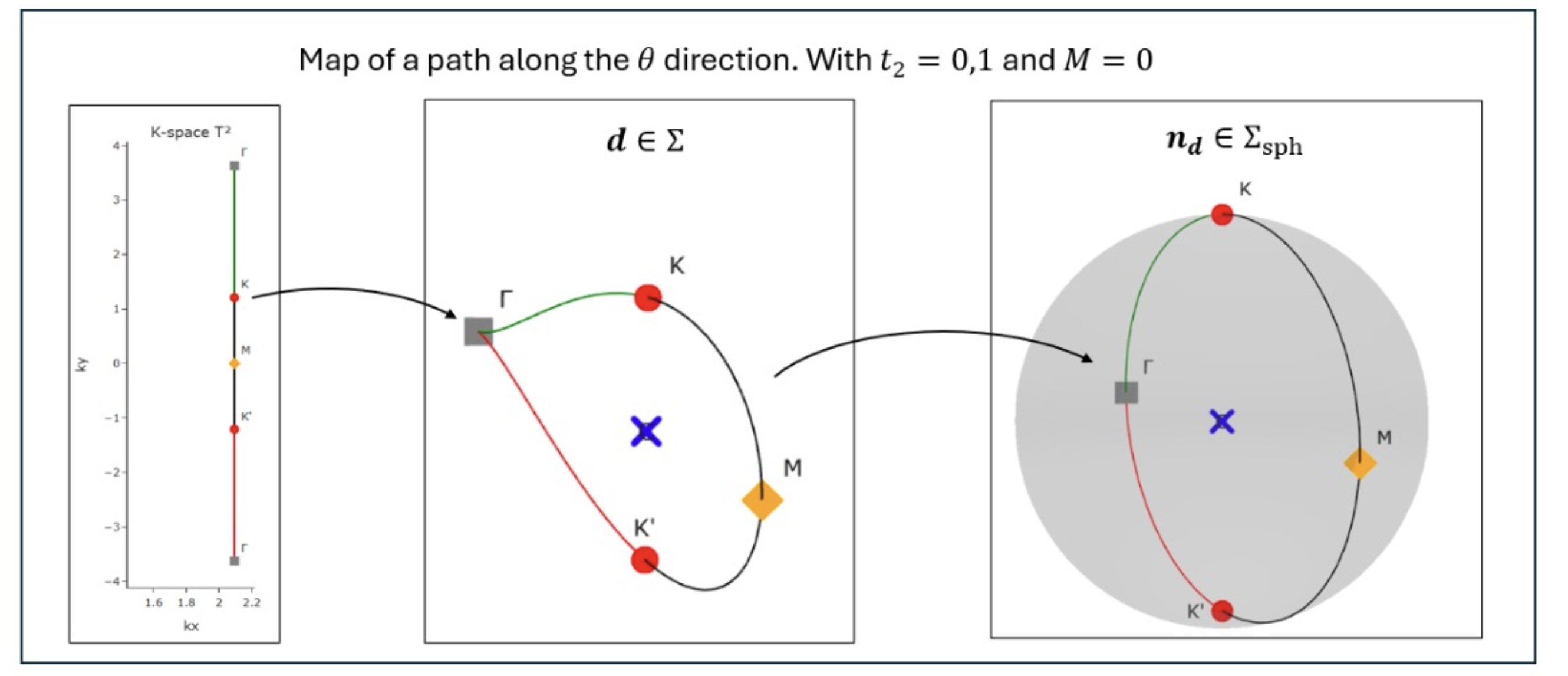} 
\vskip -0.3cm
\caption{(Top) Honeycomb lattice and Brillouin zone. The rhomboid Brillouin zone is equivalent to the hexagonal one. Within the correspondence for $t_2=0.1$ in units of $t$, we introduce the unit vector ${\bf n}_d=\frac{{\bf d}}{|{\bf d}|}$ in Eq. (\ref{dvector}).
When the ${\bf d}$ vector covers the entire sphere then this corresponds to a topological number equal to one e.g. for $M=0$. In this case, Dirac points are located at different poles on the sphere. Within the trivial phase $M=2M_c$, both Dirac points then map onto the same pole. In that case, the magnetic field (Berry curvature) does not wrap around the origin completely. This leads to a trivial winding number.
(Bottom) Path with fixed azimuthal angle $\varphi$ and parameters $M=0$, $t_2=0.1$.}
\label{honeycomb}
\end{figure}

It is judicious to introduce the definitions 
\begin{eqnarray}
d_z &=& -\gamma({\bf k}) \nonumber \\
d_x+i d_y &=& g({\bf k}).
\end{eqnarray}
The energy eigenvalues then can be written as
\begin{eqnarray}
E_{\pm}({\bf k}) = \pm \sqrt{(M-\gamma({\bf k}))^2 + |g({\bf k})|^2}.
\end{eqnarray}
For completeness, we illustrate the map from the Brillouin zone, equivalently drawn as a rhomboid, onto the sphere in Fig. \ref{honeycomb} that also clarifies the quantum phase transition from a simple geometrical view, i.e. as an equivalent geometry on the sphere encircling or not the origin of the Brillouin zone. Within the topological phase the $K$ Dirac point in red is at north pole at $\theta=0$ and the $K'$ Dirac point in red is at south pole at $\theta=\pi$. 

Close to the Dirac points, for the analytical evaluations it is useful to introduce \cite{KLHReview}
\begin{eqnarray}
\label{dvectormap}
d_x &=& -\hbar v_F |{\bf p}| \cos(\phi_p) \nonumber \\
d_y &=& -\hbar v_F |{\bf p}| \sin(\zeta\phi_p) \nonumber \\
d_z &=& \zeta 3\sqrt{3} t_2 +M.
\end{eqnarray}
For the analytical evaluations, we assume a parabolic dispersion close to $K$ and $K'$ which is satisfied for not too large values of $t_2$ (e.g. $t_2\sim 0.1t$) \cite{StephanKaryn}.
Here, $\phi_p$ represents the polar angle around each Dirac point which satisfies
\begin{equation}
\label{azimuthal}
\varphi = \zeta \phi_p \pm \pi,
\end{equation}
where $\varphi$ is the azimuthal angle on the sphere. Around the two Dirac points in the Brillouin zone, compared to the sphere definition, 
the polar angles are then introduced to rotate in different directions i.e. $\zeta=+1$ at the $K$ Dirac point and $\zeta=-1$
at the $K'$ Dirac point. Close to the Dirac points, the components $d_x$ and $d_y$ precisely give rise to the massless Dirac equation through the Hamiltonian
\begin{equation}
H^{\zeta}({\bf p}) = \hbar v_F (p_x \sigma_x + \zeta p_y \sigma_y),
\end{equation}
with $v_F=\frac{3}{2\hbar}ta$ the Fermi velocity of graphene \cite{Wallace}. We also have the correspondence between the dressed polar angle $\tilde{\theta}$ on the sphere and the deviation from each Dirac point through the wavevector ${\bf p}$:
\begin{equation}
\label{thetatilde}
\tan\tilde{\theta} = \frac{\hbar v_F|{\bf p}|}{\zeta 3\sqrt{3} t_2+M}.
\end{equation}

The many-body ground state corresponds to the lowest-energy band being filled 
\begin{equation}
|\Psi\rangle = \prod_{{\bf k}\in FBZ} (\alpha({\bf k})c^{\dagger}_{{\bf k},A} + \beta({\bf k}) c^{\dagger}_{{\bf k},B}) |0\rangle
\end{equation}
where $FBZ$ refers to the first Brillouin zone and the electron operators act on $A$ and $B$ sublattices respectively. It is associated to the state $|\psi\rangle$ on the sphere.

\begin{figure}[t]
\includegraphics[width=1\textwidth]{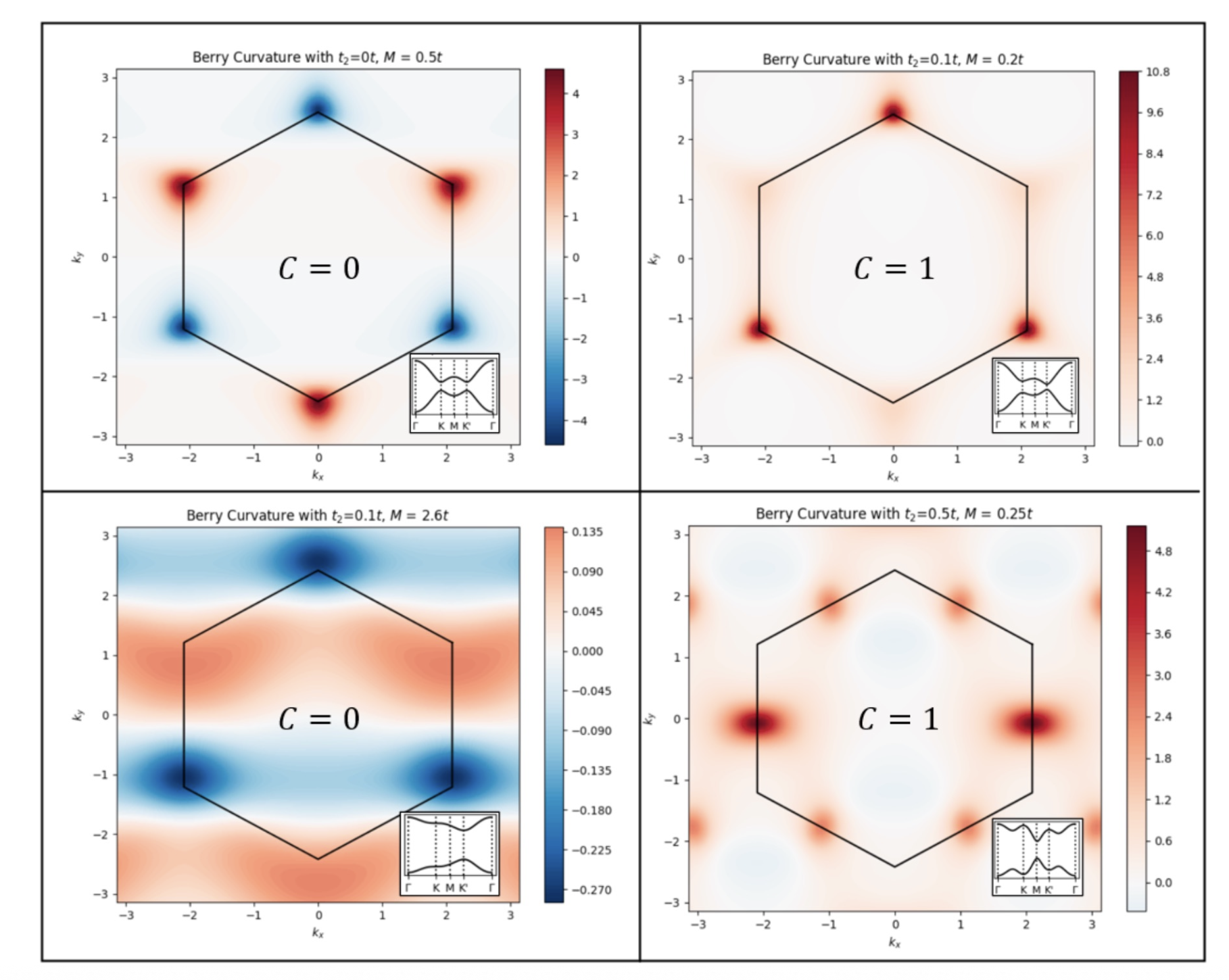} 
\vskip -0.3cm
\caption{Berry curvatures for different values of $M$ and $t_2$. (Top Left) This corresponds to the physical situation of graphene with a charge density wave substrate leading to Eq. (\ref{Berryfunction}) \cite{Eva}. (Top Right) We are within the topological phase with $C=1$ and the numbers agree with Eq. (\ref{Berryfunction}) and also with numerical results in the literature \cite{Anton}. (Bottom Left) Results within the non-topological phase with $C=0$. (Bottom Right) When increasing $t_2$ within the topological phase, there is a deviation from the Dirac approximation for the $d_x$ and $d_y$ components and the local Berry curvatures become (very) small.  Interestingly the formula $C = A_{\varphi}({\bf K}') - A_{\varphi}({\bf K})=+1$ from the Dirac points which is
exact yet works well.}
\label{Berrygraphs}
\end{figure}

Related to Figs. \ref{anglecorrespondence} and \ref{honeycomb}, the $K$ Dirac point will correspond to the polar angle $\tilde{\theta}(\theta=0)=0$ and the $K'$ Dirac
point will then correspond to the polar angle $\tilde{\theta}(\theta=\pi)$. The global topological invariant can then be defined locally as \cite{KLHReview,FractionalArticle}
\begin{equation}
\label{localmarkerlattice}
C = A_{\varphi}({\bf K}') - A_{\varphi}({\bf K}),
\end{equation}
with 
\begin{equation}
A_{\varphi}({\bf K}^{\zeta}) = - i\zeta \langle \psi_+| \partial_{\phi_p}|\psi_+\rangle.
\end{equation}
It corresponds to the addition of Berry phases around each Dirac point where the eigenstates around these two points associated to the lowest-energy band are introduced with the same $\varphi$-coherent gauge according to Eq. (\ref{azimuthal}).
The local gauge potential can be efficiently addressed numerically at the Dirac points through the Taylor formula
\begin{equation}
A_{\varphi}({\bf K}^{\zeta}) = -i \zeta \hbox{lim}_{\tilde{\theta}\rightarrow 0\ \hbox{or}\ \tilde{\theta}(\theta=\pi)} \hbox{lim}_{\Delta \phi_p\rightarrow 0} \frac{\langle \psi_+(\tilde{\theta},\phi_p)|\psi_+(\tilde{\theta},\phi_p +\Delta \phi_p)\rangle -1}{\Delta \phi_p},
\end{equation}
where $\zeta$ is $\pm$ at the $K$ and $K'$ Dirac points respectively. The formula (\ref{localmarkerlattice}) is very efficient numerically and it reproduces well the quantum phase transition at $M=M_c=3\sqrt{3} t_2$
through the jump of the topological invariant locally resolved at the Dirac points. As illustrated in Fig. \ref{Berrygraphs}, this formula works well even when the Berry curvatures become very small at $K$ and $K'$. 

At the topological quantum phase transition, the gap is closing at one Dirac point and the half topological invariant can be seen as a $\pi$ Berry phase or $\pi$ winding number related to the massive Dirac point.
From the discussion on the sphere in Sec. \ref{Monopole}, this formula $C = A_{\varphi}({\bf K}') - A_{\varphi}({\bf K})$ is gauge invariant. We emphasize here that the situation of half topological numbers can be then generalized 
to a region of the parameters space (e.g. a line) associated to the quantum anomalous Hall semimetal \cite{FractionalArticle,KarynSariah1,KarynSariah2}. This formalism builds a correspondence with the physics of surface states of three-dimensional
insulators \cite{FuKane,SekineNomura,Zhang} which also reveal a massive topological Dirac point with a $\pi$ winding number or half topological invariant (see also Section 4.6 in Ref. \cite{KLHReview}). We mention here recent efforts to observe a half-quantized Hall conductance in semimagnetic topological insulator bilayers \cite{Mogi} and in ultra-cold atoms \cite{Nascimbene} related to the $C=1/2$ parity anomaly \cite{Haldane}. A recent work also reports a half-quantized chiral edge current \cite{Zhuo}.

We also show another useful aspect of this local representation of topological properties through the Berry curvature itself. Swapping from the sphere to the Brillouin zone, it is then possible to evaluate analytically the Berry curvature assuming the linear Dirac approximation for the $d_x$ and $d_y$ components \cite{KarynLight}. Including the presence of the term $M$ identical to a Semenoff mass, we generalize the formula found in Ref. \cite{KarynLight} by one of us as
\begin{eqnarray}
\label{Berryfunction}
F_{p_x,p_y}^{\zeta} &=& -\frac{1}{2(M + \zeta 3\sqrt{3}t_2)^2} \hbox{Im} \left( \langle \psi_+|\partial_{p_x} H^{\zeta} |\psi_-\rangle \langle \psi_-|\partial_{p_y} H^{\zeta} |\psi_+\rangle \right) \nonumber \\
&=& \zeta \frac{\hbar^2 v_F^2}{2(M + \zeta 3\sqrt{3}t_2)^2}\cos\tilde{\theta},
\end{eqnarray}
where 
\begin{equation}
\cos\tilde{\theta} = \langle \psi_+| \sigma_z | \psi_+\rangle = -2 A_{\varphi}(\tilde{\theta}).
\end{equation}
We emphasize that $|\psi_+\rangle$ (called $|\psi \rangle$ in the preceding Section) and $|\psi_-\rangle$ correspond to the {\it lowest-energy} and {\it upper-energy} eigenstates of the spin-1/2 particle 
\begin{equation}
|\psi_-\rangle = -|\beta| e^{-i \frac{\varphi}{2}}|+\rangle + |\alpha| e^{i \frac{\varphi}{2}}|-\rangle.
\end{equation}
These two eigenstates are then related to the lowest and upper energy bands in the Haldane model.
A similar relation between Berry curvature and pseudo-spin response was derived in relation to the Thomas precession and Dirac equation in 1991 and 1994 \cite{Mathur,MathurShankar}. 
This leads to the results of Fig. \ref{Berrygraphs}. We verify the height of Berry curvature surrounding each Dirac point with the topological phase e.g. for $M=0$
\begin{eqnarray}
F_{p_x,p_y}^{\zeta} = -\zeta \frac{\hbar^2}{24 t_2^2}(t a)^2 (2A_{\varphi}(\tilde{\theta})).
\end{eqnarray}
When summing the Berry curvature at the two Dirac points then this equally defines the topological invariant. In this sense, the quantum Hall conductivity written in terms of Berry curvatures \cite{TKNN} can also be resolved 
locally from the Dirac points modulo a prefactor that is also useful to obtain an estimate of the ratio $\frac{t_2}{t}$. We assume here that $t_2\neq 0$ because Eq. (\ref{Aphi}) is applicable in the presence of a radial
magnetic field on the sphere implying then a term $t_2$ on the lattice.

In Sec. \ref{light}, the response to circularly polarized light, i.e. through the photo-induced currents, will precisely allow us to resolve this information from the Dirac points \cite{PhilippAdolfoKaryn}. We will illustrate an application including the mass 
term $M$. 

\subsection{Populations and Pseudo-Spin Response through a Local Marker in Momentum Space}

\begin{figure}[t]
\includegraphics[width=1\textwidth]{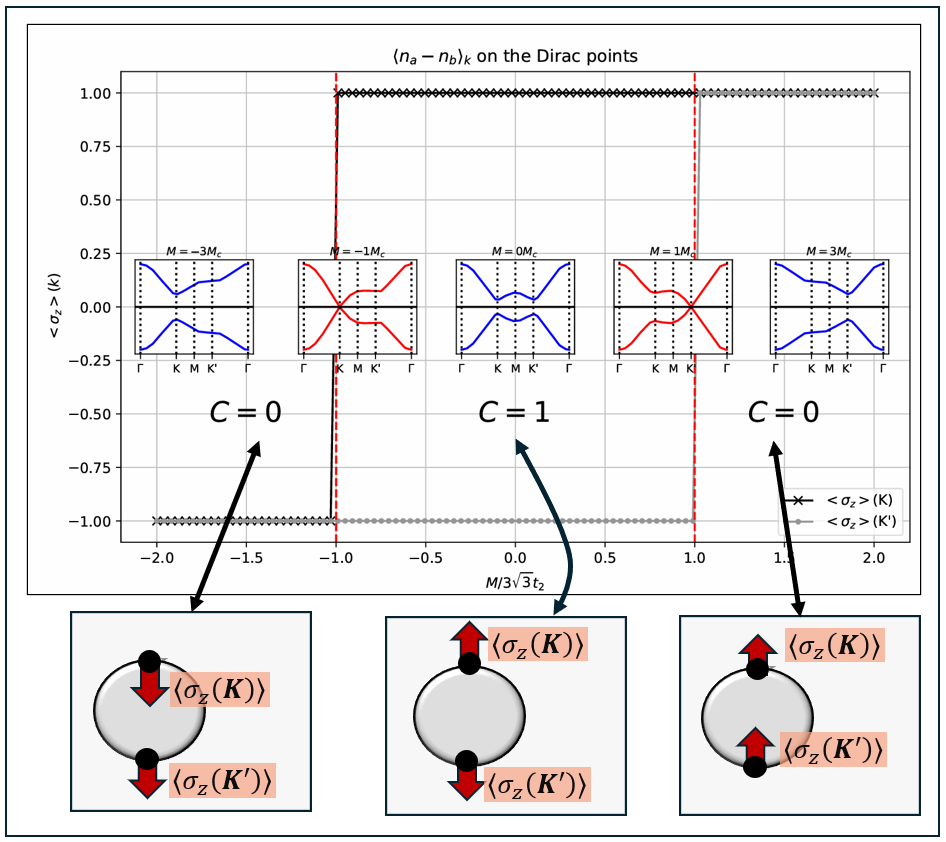} 
\vskip -0.3cm
\caption{Pseudo-spin response at the two Dirac points and illustration of the phase
transition through the flip of the pseudo-spin at one pole. $\langle \sigma_z(\bm k)\rangle$ evaluated at the Dirac points as a function of $\frac{M}{3\sqrt 3 t_2}$. 
The top figure shows the five energy diagrams along the Brillouin zone axis. The black line response is relative to the $K$-point, which is mapped to the north pole of the sphere, while the grey one is relative to the $K'$-point, which corresponds to the south pole. As long as $|M|<3\sqrt 3 t_2$, we are in the topological phase and so the pseudospin responses at $K$ and $K'$ are in opposite directions. In the trivial phase the pseudospins point in the same directions. This corresponds to the topological band inversion.}
\label{populations}
\end{figure}

It is then useful to build a precise correspondence between populations on sublattices $A$ and $B$ and the pseudo-spin response along $z$ direction within the ground state.
From the correspondence on the sphere, 
\begin{equation}
C = \frac{\cos\tilde{\theta}({\bf K}) - \cos\tilde{\theta}({\bf K}')}{2} = \frac{\langle \sigma_z({\bf K}) \rangle - \langle \sigma_z({\bf K}') \rangle}{2}.
\end{equation}
We illustrate the usefulness of this local marker in momentum space associated to the pseudo-spin response in Fig. \ref{populations}.
The populations on $A$ and $B$ sublattices with e.g. $n_{{\bf k},A}=c^{\dagger}_{{\bf k},A} c_{{\bf k},A}$ take the forms
\begin{eqnarray}
\label{density}
\langle n_{{\bf k},A}\rangle &=& \langle \Psi |n_{{\bf k},A}|\Psi\rangle = |\alpha({\bf k})|^2 = \frac{1}{2} {\color{blue}-} \frac{\gamma({\bf k})-M}{2|{\bf d}({\bf k})|} \\ \nonumber
\langle n_{{\bf k},B}\rangle &=& \langle \Psi|n_{{\bf k},B}|\Psi\rangle = |\beta({\bf k})|^2 = \frac{1}{2} + \frac{\gamma({\bf k})-M}{2|{\bf d}({\bf k})|}.
\end{eqnarray}
Close to the Dirac points, since $g({\bf k})\rightarrow 0$, then we obtain the simple general form 
\begin{equation}
\langle \sigma_z({\bf K}^{\zeta})\rangle = |\alpha({\bf K}^{\zeta})|^2-|\beta({\bf K}^{\zeta})|^2 = -\frac{\gamma({\bf K}^{\zeta}) - M}{|\gamma({\bf K}^{\zeta})-M|}=sgn(M+\zeta 3\sqrt{3} t_2).
\end{equation}
It is also possible to re-interpret the $sgn$ function at each Dirac point through a mass sign in Eq. (\ref{dz}) which is then dressed with the $M$ term \cite{Haldane,Orsay,Nathan}. If we relate the population in momentum space with the photoluminescence intensity in Ref. \cite{C2N}, then Eqs. (\ref{density}) may be resolved through the Stokes parameters. In that article, the authors then introduce the valley Chern number \cite{Haldane}. Information about pseudo-spin physics of Skyrmions may also be accessed
through time of flight in cold atoms \cite{Alba}.

We also emphasize on the fact that the in-plane pseudo spin observable in momentum space acquires an important representation on the sphere related to the local Berry curvature $F_{\tilde{\theta}\varphi}$:
\begin{equation}
|\langle \psi| \sigma_x |\psi\rangle| = |\alpha^*({\bf k})\beta({\bf k}) + \alpha({\bf k})\beta^*({\bf k})| = \sin\tilde{\theta}.
\end{equation}
The in-plane pseudo-spin response for the Haldane model was recently addressed in Ref. \cite{IkedaRayan} associated to sublattice coherence effects.

\subsection{Real-Space Analysis and Comparison with the Effective Magnetic Moment of the Monopole}
\label{realspace}

From Parseval-Plancherel theorem, the real space pseudo-spin response reads
\begin{equation}
\langle \sigma_z({\bf R}_i)\rangle = -\frac{1}{N} \sum_{\bf k} \frac{\gamma({\bf k})-M}{|{\bf d}({\bf k})|},
\end{equation}
For $M=0$, the function $\gamma$ is odd under the transformation $k_y\rightarrow -k_y$ such that $\langle \sigma_z({\bf R}_i)\rangle=0$. 

\begin{figure}[t]
\includegraphics[width=1\textwidth]{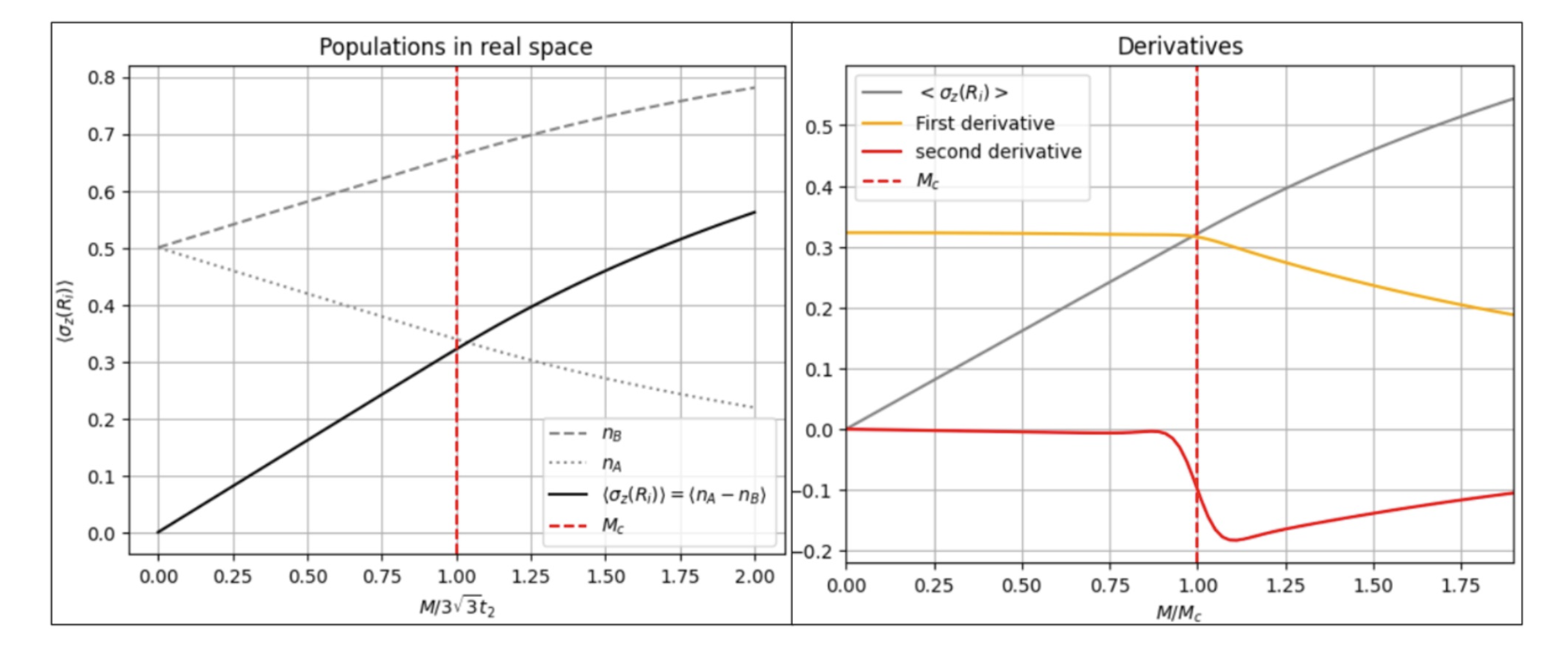} 
\vskip -0.3cm
\caption{Numerical study of the populations in real space and of $\langle \sigma_z({\bf R}_i)\rangle$ as a function of $M$ with $t_2=0.1t$ as in Fig. \ref{honeycomb}.
First and second derivatives as a function of $M$. Here, we remind that $M_c=3\sqrt{3}t_2$ which is analogous to $B$ for the monopole. The calculation is done by doing a discrete Fourier transform in a
$100 \times 100$ finite size lattice in Periodic Boundary Conditions.}
\label{pseudospin}
\end{figure}

In Fig. \ref{pseudospin}, we analyse the properties of this local marker in real space comparatively to $\kappa$ in Eq. (\ref{kappaspin}).
When varying the parameter $M$, corresponding e.g. to describe the effects of a charge density wave substrate \cite{Eva}, we do reveal a linear response behavior for the pseudo-spin locally on the lattice until the phase transition similar to the analysis of the response of the monopole. The first derivative in {\it yellow} of the pseudo-spin is also fixed until the topological phase transition. We report a similar behavior as the susceptibility $\chi=B\frac{\partial \kappa}{\partial M}$ in Eq. (\ref{chitopo}). The first derivative from the real space analysis on the lattice in Fig. \ref{pseudospin} is around $\sim 0.33...$, which is approximately half of $\chi$. 
The susceptibility in yellow in Fig. \ref{pseudospin} clearly distinguishes the position (location) of the transition. In Fig. \ref{pseudospin}, we show in {\it red} the second derivative 
of  $\langle \sigma_z({\bf R}_i)\rangle$ as a function of $\frac{M}{M_c}$.

The comparison between the local marker in real space $\langle \sigma_z({\bf R}_i)\rangle$ and the theory of the monopole's spin response is interesting since in general for any observable a careful correspondence between quantities evaluated on the lattice and on the sphere is required. For the quantum Hall response it is possible to re-write the global topological invariant in the plane as an evaluation of the Berry curvature on the sphere through $F_{\theta\varphi}$ with a measure of the area integration as $d\theta d\varphi$ \cite{KLHReview}. There is also a simple understanding of this mapping from the Kubo formula for the quantum Hall response \cite{KLHReview}.
For the pseudo-spin response, this is evaluated on the spherical surface and we have verified that making a flat-space simplification does not reproduce the same results in particular at the transition.
The Jacobian of a transformation necessitates some thoughts in general. As long as Eqs. (\ref{dvectormap}) are satisfied i.e. with an energy spectrum of the form $\pm \sqrt{(\hbar v_F |{\bf p}|)^2 + (\zeta 3\sqrt{3}t_2+M)^2}$ around each Dirac point we can justify the results as follows. On the lattice, we evaluate 
\begin{equation}
\frac{1}{(2\pi)^2}\iint \langle \sigma_z({\bf k})\rangle dk_x dk_y.
\end{equation}
In the vicinity of the north pole or of the $K$ Dirac point, from Eq. (\ref{thetatilde}), we can develop $dk_x dk_y\rightarrow (2\pi) |{\bf p}| d|{\bf p}|$ such that $(\hbar v_F)^2 d|{\bf p}||{\bf p}|=d\theta 
\sin \theta M_c^2$ with $M_c=3\sqrt{3}t_2$. We develop the area of the rhomboid Brillouin zone from a disk centered around the $K$ Dirac point. In this correspondence the radius of the sphere is $M_c$ whereas the radius of the disk around the $K$ Dirac point is $\hbar v_F$. 
To reproduce the lattice  result from the sphere which shows a linear relation between $\theta$ and  $|{\bf p}|$ around north pole then this requires e.g. to rescale 
 $\hbar v_F=1$ and $M_c=1$ corresponding to $t_2\sim 0.19$ and $t=\frac{2}{3}$ when setting the lattice spacing to unity. In this way,
\begin{equation}
\frac{1}{(2\pi)^2}\iint \langle \sigma_z({\bf k})\rangle dk_x dk_y \sim \frac{1}{\pi}\kappa \sim 0.21\frac{M}{M_c}.
\end{equation}
This approximation tends to provide a good justification of why $\langle \sigma_z({\bf R}_i)\rangle$ is related to $\kappa$ such that it also shows a plateau within the topological phase, as observed numerically  in Fig. \ref{pseudospin}.
For larger values of $t_2$, numerically we observe some curvature effects on the plateau for $\langle \sigma_z({\bf R}_i)\rangle$ which can be associated to deviations from the parabolic band approximation
and to the flattening of the bands \cite{StephanKaryn}. The topological phase transition remains visible. Yet, the topological marker introduced on the sphere in Fig. \ref{spinresponse} remains applicable for all values of $B$ or $t_2$ on the lattice.

It is then useful to present the numerical analysis in real space, resolved in each sublattice, associated to the Fourier transforms of $\langle n_{{\bf k},A}\rangle$ and $\langle n_{{\bf k},B}\rangle$:
\begin{eqnarray}
\langle n_{{\bf k},A} \rangle = |\alpha({\bf k})|^2 &=& \sum_{\bf R} e^{i {\bf k}\cdot {\bf R}} C({\bf R},A) \nonumber \\
\langle n_{{\bf k},B}\rangle =|\beta({\bf k})|^2 &=& \sum_{\bf R} e^{i {\bf k}\cdot {\bf R}} C({\bf R},B).
\end{eqnarray}
In this way, the topological invariant takes the equivalent form 
\begin{equation}
\label{invariantcorrelation}
C = \frac{1}{2} \sum_{\bf R} (C({\bf R},A) - C({\bf R},B))\left(e^{i{\bf K}\cdot {\bf R}} - e^{i{\bf K}'\cdot {\bf R}}\right).
\end{equation}
This can also be re-written as
\begin{equation}
C = i \sum_{\bf R}(C({\bf R},A) - C({\bf R},B)) e^{i K_x R_x} \sin(K_y R_y).
\end{equation}
For one-dimensional topological systems, it is in general possible to evaluate the summation analytically \cite{FrederickLoicKaryn,FrederickLoicOlesiaKaryn,KarynFanMagali}.
It is then useful to introduce the correlation function in real space associated to the topological invariant
\begin{equation}
\label{f}
f({\bf R}) = C({\bf R},A) - C({\bf R},B) = \frac{1}{N}\sum_{\bf q} e^{-i{\bf q}\cdot {\bf R}} \langle \sigma_z({\bf q})\rangle = \frac{1}{N}\sum_{\bf q} e^{-i{\bf q}\cdot {\bf R}} \left(\frac{-\gamma({\bf q})+M}{\sqrt{(\gamma({\bf q})-M)^2 + |g({\bf q})|^2}}\right).
\end{equation}
When $M\gg M_c=3\sqrt{3}t_2$, $\langle \sigma_z({\bf q})\rangle \rightarrow +1$ and therefore correlation functions are very short-range such that $f({\bf R})\rightarrow \delta({\bf R})$ traducing indeed an {\it insulating phase} such that $C=0$
due to the presence of the sine function in the sum. Within the topological phase, the function $e^{iK_x R_x}$ is real for nearest neighbors of the same flavour. Therefore, this requires to have longer-range of correlations. Yet, the phase shows a
gap such that a relatively short decay also occurs. Numerically, we identify an exponential decay for $f({\bf R})$ at short distances for length scales up to 10 units of lattice vectors. Correlation functions also decay in a fast way for the Kane-Mele model \cite{StephanKaryn}. Within the {\it topological phase} e.g. for $M=0$, if we insert Eq. (\ref{f}) into Eq. (\ref{invariantcorrelation}) we can verify that $C=1$. For the p-wave Kitaev superconductor, at half-filling and for $t=\Delta$, the nearest-neighbor correlator
reproduces the topological invariant \cite{FrederickLoicOlesiaKaryn}.  At the transition, from this analysis, the correlation length diverges from the trivial phase.

\subsection{Relation to Local Responses from Circularly Polarized Light}
\label{light}

Here, we show how the pseudo-spin response and the Berry curvature are locally measured through the responses to circularly polarized light and develop the theory of the quantum phase transition.

It is useful to remind that for a two-level system the light matter interaction takes the form
\begin{equation}
\delta H_{\pm} = A_0 e^{\pm i\omega t} |+\rangle \langle -| +h.c. = A_0 e^{\pm i\omega t}\sigma^+ +h.c.
\end{equation}
The classical vector potential is described through its components $A_x=A_0\cos \omega t$ and $A_y=\mp A_0 \sin \omega t$. This is equivalent as a circularly polarized
vector potential ${\bf A}_{\pm} = A_0 e^{-i\omega t} ({\bf e}_x \mp i {\bf e}_y)= A_0 e^{-i\omega t} (\mp {\bf e}_{\varphi})$.
It is useful to introduce the inter-band transition probabilities associated to each polarization \cite{KarynLight,KLHReview}
\begin{equation}
\Gamma_{\pm} = \frac{2\pi}{\hbar} |\langle \psi_+| \delta H_{\pm} |\psi_-\rangle|^2 \delta (E_b-E_a \mp \hbar \omega).
\end{equation}
The meaning of right $(+)$ and left-handed $(-)$ polarizations can be understood from the {\it resonance} which is obtained through the transformation 
$|-\rangle \rightarrow e^{\mp \frac{i\omega t}{2}}|-\rangle$ and $|+\rangle \rightarrow e^{\pm \frac{i\omega t}{2}}|+\rangle$. 
On the lattice $|+\rangle$ refers to
the $|A\rangle$ sublattice polarization and $|-\rangle$ refers to the $|B\rangle$ sublattice polarization. 
In this way, the time-dependent phases can be interpreted as an energy shift 
$E_b - E_a = \pm \hbar\omega$. 

From the form of the eigenstates $|\psi_+(\tilde{\theta},\varphi)\rangle=|\psi_+\rangle$ and $|\psi_-(\tilde{\theta},\varphi)\rangle=|\psi_-\rangle$
 in the presence of the mass $M$, we can then generalize the geometrical function $\alpha(\theta)$ introduced in previous articles \cite{KarynLight,KLHReview} as:
 \begin{equation}
\alpha(\tilde{\theta}) =  |\langle \psi_+| \sigma_x|\psi_-\rangle|^2 +  |\langle \psi_+| \sigma_y|\psi_-\rangle|^2 = \sin^4\frac{\tilde{\theta}}{2} + \cos^4\frac{\tilde{\theta}}{2}
= \alpha(\pi-\tilde{\theta}).
\end{equation}
For $M=0$, one of us has shown that this function at the two Dirac points is precisely related to the square of the topological invariant \cite{KarynLight,KLHReview}.
Close to $\tilde{\theta}=0$ we can equivalently write
\begin{equation}
\sin^4\frac{\tilde{\theta}}{2} + \cos^4\frac{\tilde{\theta}}{2} = \langle \sigma_z(\theta=0)\rangle 
\end{equation}
and close to $\tilde{\theta}(\theta=\pi)$, from definitions in Appendix A, we also have
\begin{equation}
\sin^4\frac{\tilde{\theta}}{2} + \cos^4\frac{\tilde{\theta}}{2} = \langle \sigma_z(\tilde{\theta}(\theta=\pi))\rangle +2C^2.
\end{equation}
As long as we are within the topological phase, the points $\tilde{\theta}=0$ and $\tilde{\theta}=\pi$ correspond respectively to the points $\theta=0$ and $\theta=\pi$. 
The functions $\sin^4\frac{\tilde{\theta}}{2}$ and $\cos^4\frac{\tilde{\theta}}{2}$ can also be interpreted in terms of the topological responses $(A_{\varphi}'(\theta_c^-))^2$ and $(A_{\varphi}'(\theta_c^+))^2$; see Appendix \ref{AppendixA}.
The transition probabilities at $\theta=0$ from the right-handed light source ($\Gamma_{+}$) and at $\theta=\pi$ from the left-handed light source ($\Gamma_{-}$) reveal quantized responses 
$\Gamma_{+}(0) = \Gamma_{-}(\pi)$. The quantized light information at resonance then reveals the definition of the topological invariant resolved at the Dirac points.
Fixing $\tilde{\theta}$, when varying $\omega$ we can then resolve the form factor associated to each light response. When crossing the phase transition,
the two poles become equivalent from the eigenstates structure. In that case, $\Gamma_{+}(0)=\Gamma_{+}(\pi)$ from the energy conservation. The light responses also reveal the pseudo-spin state at the two Dirac 
points. 

Related to the left- and right-handed waves on the sphere, in the plane this turns into
\begin{equation}
\zeta \phi_p \pm \pi = \pm \omega t = \varphi
\end{equation}
Compared to the definitions on the sphere, this implies $\Gamma_{+}(\pi)\rightarrow \Gamma_{LP}({\bf K}')$
and $\Gamma_{-}(\pi)\rightarrow \Gamma_{RP}({\bf K}')$ where $LP$ and $RP$ designate circular light and right polarizations in the plane. Then, we verify the topological structure ${\Gamma}_{RP}({\bf K})={\Gamma}_{RP}({\bf K}')$ within the topological phase and ${\Gamma}_{RP}({\bf K})={\Gamma}_{LP}({\bf K}')$ within the trivial phase. For these responses,  we have $A_0=\frac{\mathcal{E}}{\Delta({\bf k})}$ where $\Delta({\bf k})$ corresponds to the resonance energy gap for a wavevector ${\bf k}$ and 
$\mathcal{E}$ is the electric field.  At the transition, the gap is closing at the $K'$ point which means effectively that ${\Gamma}_{RP}({\bf K}')={\Gamma}_{LP}({\bf K}')=0$ if we assume a light source with $\omega\neq 0$.

\begin{figure}[t]
\includegraphics[width=1.05\textwidth]{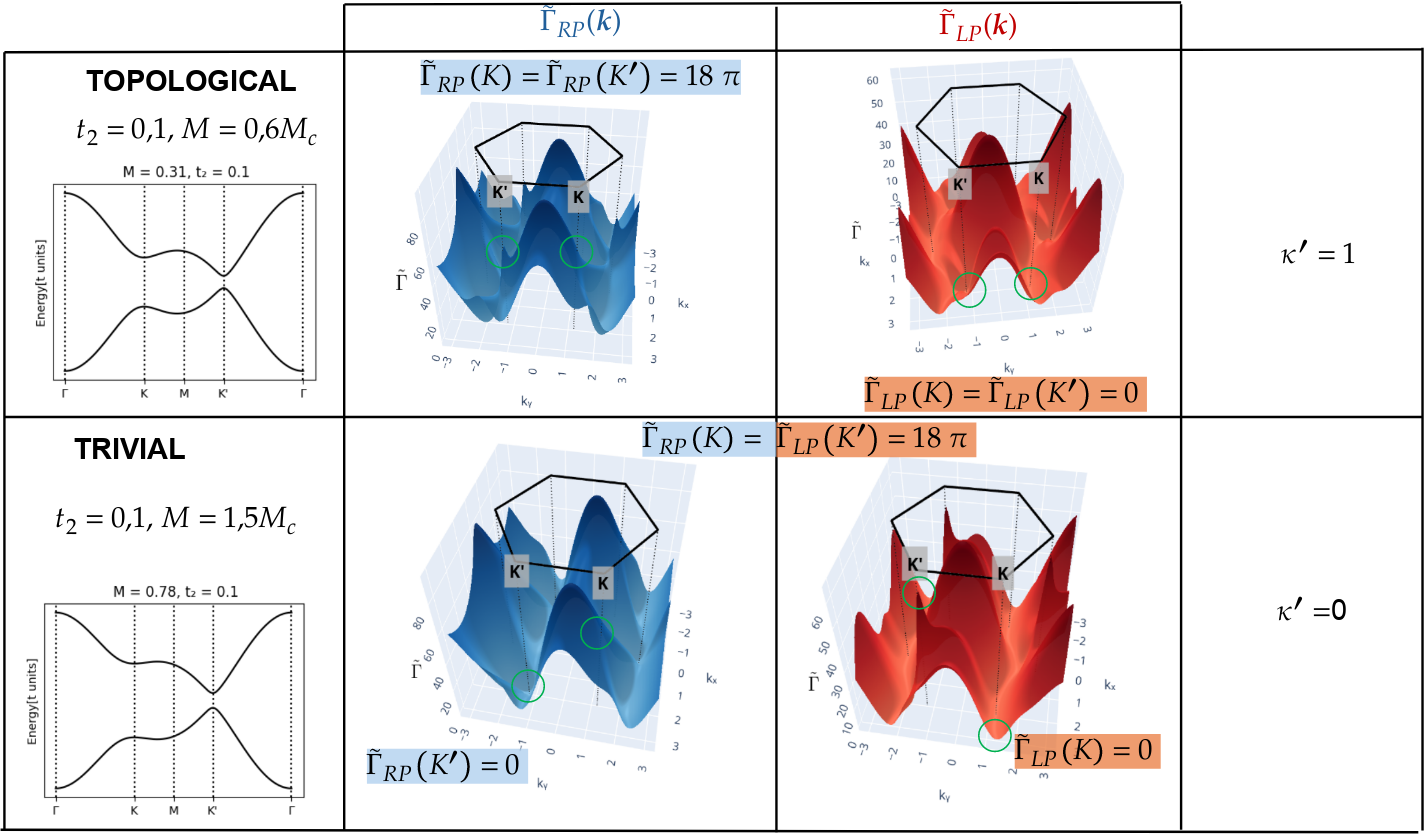} 
\vskip -0.3cm
\caption{Light responses for the right- and left circular polarizations within the topological phase and the trivial phase for $t_2=0,1$ with $t=1$. 
We evaluate numerically the photo-induced currents and verify the analytical formulas for each circular drive $\tilde{\Gamma}_{RP}$ and $\tilde{\Gamma}_{LP}$ at the two Dirac points. Within the topological phase the two Dirac points interact with the right-handed circular drive and we verify that the height of the signal peaks remains invariant when varying $M$. We verify the value of $\kappa'$ in the two phases.
At the phase transition, the gap is closing at the $K'$ Dirac point, i.e. the light response with frequency $\omega\neq 0$ occurs only at the $K$ Dirac point. The $\frac{1}{2}$ light response can then define the $\frac{1}{2}$ invariant at the topological phase
transition. Above the transition, the $K'$ Dirac point interacts with the left-handed circular drive and the $K$ Dirac point with the right-handed circular wave.}
\label{LightResponse}
\end{figure}

We emphasize on the fact that the detection of the topological invariant through circularly polarized light attracts some attention in the community related to the quantum Hall conductivity \cite{NathanPeter,DFG,Japan}. The light signal is then related
to the integral of the Berry curvature when summing the photo-induced currents on all momenta in the Brillouin zone. In Ref. \cite{PhilippAdolfoKaryn}, we have suggested that the same photo-induced response written in terms of Berry curvatures may be revealed from the Dirac points. Here, we elaborate on this fact and show that the method from the Dirac points is very efficient quantitatively to detect the quantum phase transition. The photo-induced response for the {\it currents} takes the form \cite{NathanPeter,PhilippAdolfoKaryn},
\begin{equation}
\Gamma_{\lambda}(\omega,{\bf k}) = \frac{2\pi}{\hbar}\left(\frac{{\mathcal E}}{\Delta({\bf k})}\right)^2 \left|\left\langle u_{\bf k}\left | \left(\lambda i \frac{\partial H}{\partial p_x}+\frac{\partial H}{\partial p_y}\right)\right|l_{\bf k}\right\rangle\right|^2 \delta(\hbar\omega-\Delta({\bf k})).
\end{equation}
In this formula the lowest and upper energy eigenstates are introduced as $|l_{\bf k}\rangle$ and  $|u_{\bf k}\rangle$, $\lambda$ refers to a right or left circular polarization corresponding to $\lambda=\pm$ respectively.
In Sec. \ref{Map}, $p_x$ and $p_y$ are introduced as wavevector components measured from each Dirac point. 
These responses are in accordance with the transition probabilities introduced on the sphere and the selection rules mentioned above. 
It is then useful e.g. to integrate the responses in frequency. In experiments, for $t_2\sim 0.1t$, this is equivalent to fix $\hbar\omega=\Delta({\bf K})$ and integrate on momenta (see Fig. \ref{LightResponse}) and similarly
for the response at the other Dirac point obtained when fixing the resonance $\hbar\omega=\Delta({\bf K}')$.
We can then verify the precise result
\begin{equation}
{\Gamma}_{RP}({\bf K})-{\Gamma}_{LP}({\bf K})= \frac{2\pi}{\hbar}\frac{{\mathcal E}^2}{\Delta({\bf K})^2} \langle u_{\bf K} | i \partial_{p_x} H | l_{\bf K}\rangle
\langle l_{\bf K} | \partial_{p_y} H | u_{\bf K}\rangle = \frac{2\pi}{\hbar} {\mathcal E}^2 F_{p_x p_y}({\bf K})
\end{equation}
with $F_{p_x p_y}$ the Berry curvatures evaluated in Sec. \ref{Map} in the presence of $M$. We also have
\begin{equation}
{\Gamma}_{RP}({\bf K}')-{\Gamma}_{LP}({\bf K}')= \frac{2\pi}{\hbar}\frac{\mathcal{E}^2}{\Delta({\bf K}')^2} \langle u_{{\bf K}' }| i \partial_{p_x} H | l_{{\bf K}'}\rangle
\langle l_{{\bf K}'} | \partial_{p_y} H | u_{{\bf K}'}\rangle = \frac{2\pi}{\hbar} {\mathcal E}^2 F_{p_x p_y}({\bf K}').
\end{equation}
If we add (sum) these two equations multiplying by $\Delta({\bf k})^2=4E_-(\tilde{\theta})^2$ resulting then in the responses $\tilde{\Gamma}$ locally then 
this measures the topological invariant $C$. It is then relevant to introduce
\begin{equation}
\label{kappa}
\kappa'=\frac{\tilde{\Gamma}_{RP}({\bf K}) + \tilde{\Gamma}_{RP}({\bf K}') - \tilde{\Gamma}_{LP}({\bf K}) - \tilde{\Gamma}_{LP}({\bf K}')}{\tilde{\Gamma}_{RP}({\bf K}) + \tilde{\Gamma}_{RP}({\bf K}') +\tilde{\Gamma}_{LP}({\bf K}) + \tilde{\Gamma}_{LP}({\bf K}')}=C,
\end{equation}
with
\begin{equation}
\tilde{\Gamma}_{RP}({\bf K}) + \tilde{\Gamma}_{RP}({\bf K}') +\tilde{\Gamma}_{LP}({\bf K}) + \tilde{\Gamma}_{LP}({\bf K}') = \frac{2\pi}{\hbar} {\mathcal E}^2 (\hbar v_F)^2.
\end{equation}

We summarize the results in Fig. \ref{LightResponse}. Within the topological phase, the right-handed circular drive induces two peaks at $K$ and $K'$ of height $18\pi\approx 56,55$ if we set $t_2=0.1$, $t=1=\mathcal{E}$ and similarly for the lattice spacing.
We emphasize on the fact that the heights of these peaks remain identical within the topological phase. When crossing the phase transition, we see clearly that the $K'$ Dirac point now interacts with the left-handed circular drive and the peak height remains the same. From the peak heights resolved at the two Dirac points it is now possible to read if the phase is topological or not. It also encodes quantitative information on local Berry curvature and pseudo-spin response. 

Locating the transition with the topological invariant $C_{1/2}=\frac{1}{2}$ then corresponds to detect half of the signal from the two Dirac points compared to the topological phase (i.e. only one peak with the same height or intensity 
located at the $K$ Dirac point), with the right-handed circularly polarized light. A half signal is also measurable for the quantum anomalous Hall semimetal \cite{KarynSariah2} and quantum spin Hall semimetal \cite{KLHQSHSemimetal} in this way.

This analysis can be generalized to the quantum spin Hall effect \cite{KarynLight}. When coupling a plane of graphene with a topological thin material described through a Haldane model \cite{Haldane}, we have shown the induction of a $\mathbb{Z}_2$ topological state where through proximity effect  the graphene system acquires a topological number different in sign compared to the one in the Haldane model \cite{DFG2}. This produces a $\mathbb{Z}_2$ topological number 
$C_h-C_g=\pm 2$ with e.g. $C_h=1$ and $C_g=-1$ referring to the topological numbers in the two planes. The proximity effect is induced from a AA-BB stacking where we introduce $r$ as the inter-planes hopping term. This generalizes the quantum spin Hall effect
of Kane and Mele \cite{KaneMele1,KaneMele2} to the situation with different $d_z$ components in the two planes.
The graphene is then described through a ${\bf d}_g$-vector of the form
\begin{equation}
{\bf d}_g = \left(d_x,d_y,-\frac{|r|^2}{|d_z^h({\bf k})|}d_z^h({\bf k})\right),
\end{equation}
with $d_z^h$ as in Eq. (\ref{dvectormap}) with $M=0$. This analysis shows that the quantum anomalous Hall effect with $\mathbb{Z}$ topological order \cite{Haldane}
and the quantum spin Hall effect with $\mathbb{Z}_2$ topological number \cite{KaneMele1,KaneMele2,Sheng} are both present in coupled-planes systems when going from one to two dimensions. The $\mathbb{Z}_2$ invariant
is then measurable through circularly polarized light \cite{KarynLight}: in the thin material described through a Haldane model the two Dirac points interact with the right-handed circularly polarized light whereas in the topological graphene plane
the two Dirac points interact with the left-handed circularly polarized light, following the precise protocol described above. Inverting the sign of the invariant is equivalent to invert the roles of $A$ and $B$ sublattices in the eigenstates
at the two Dirac points corresponding then to a modification of the direction of the circular drive.

\section{Application to Coupled Planes Materials through the Ramanujan Alternating Infinite Series}
\label{coupledplanes}

\subsection{Alternating $\mathbb{Z}$ and $\mathbb{Z}_2$ Topological States in Coupled Planes Models}

Here, we generalize the analysis to multi-layered systems and we are questioning the thermodynamical situation in the vertical $z$ direction perpendicular to the plane(s). We are then addressing
a situation where successive planes will show an alternating topological invariant of the form $(-1)^j$ with $j=0,1,...$. 

For three planes, this can be realized with a topological layer described through
the Haldane model in between two layers of graphene. The protocol can be generalized adding further thin layers. In principle, it is also possible to find a material where the Berry curvature or
topological invariant would alternate in sign in the $z$ direction. The system of two layers discussed above and in Ref. \cite{DFG2} was motivated from an analysis in ultra-cold atoms in optical lattices that may be generalized for multi-layered systems
with alernating topological invariants. The model can also be realized when assembling successively QAH thin materials and QSH thin materials. 
The inter-layer hopping term is smaller than the smallest energy gap in the system such that topological properties are well defined in each plane.
For a finite number of planes, we can introduce
two invariants, the quantum Hall conductivity
\begin{equation}
\sigma_{xy} = \frac{e^2}{h}\sum_{j=0}^N (-1)^j,
\end{equation}
and from the divergence theorem a $\mathbb{Z}_2$ invariant that describes the physics on top and bottom surfaces.
In physics, the divergence theorem was introduced associated to fluid mechanics by J. L. Lagrange in 1762 and was then addressed
by C. F. Gauss and M. Ostrogradsky. In mathematics, it is associated to the Green's theorem. We can then rephrase the divergence theorem in the present situation as
\begin{equation}
C_{\mathbb{Z}_2}=C_N - C_{N=0} = C_{N=0}((-1)^N -1) = \frac{1}{2\pi}\iiint \frac{\partial F_z=j}{\partial z} dz dk_x dk_y,
\end{equation}
such that the topological invariant in each plane reads
\begin{equation}
C_j = \frac{1}{2\pi}\oiint F_jdk_x dk_y  = (-1)^j 
\end{equation}
with
\begin{equation}
F_j =(-1)^j F_{k_x k_y} \delta_{zj} = F_{z=j}.
\end{equation}
Here, $F_{k_x k_y}$ is the Berry curvature in one plane that corresponds to $F_{p_x p_y}^{\zeta}$ close to each Dirac point, as described in the preceding Sections.
When $N$ is even with (N+1) planes, the situation is similar to the $\mathbb{Z}$ quantum Hall effect i.e. $C_{\mathbb{Z}_2}=0$ and $\sigma_{xy}=\frac{e^2}{h}$. When $N$ is odd with (N+1) planes, then the situation is yet
comparable to the quantum spin Hall effect i.e.  $C_{\mathbb{Z}_2}=+2$ and $\sigma_{xy}=0$. For $N$ finite, we have a {\it even-odd} effect alternating QAH and QSH states successively.

A question then comes from the infinite series of Ramanujan
\begin{equation}
{\mathcal S} = \sum_{j=0}^{\mathcal R} (-1)^j = \frac{1}{2}.
\end{equation}
The symbol ${\mathcal R}$ refers to the Ramanujan way of performing the infinite series. We are then wondering what is the physical meaning of this $\frac{1}{2}$ related to the material i.e. how should we think about the thermodynamical limit then?
From the physical point of view, one may argue that an interface between a QAH state and the vacuum or a non-trivial material allows for an interface with a jump of the topological invariant corresponding within our formulation to a situation
with $M=M_c$ implying a $\frac{1}{2}$ topological number of one massive Dirac point characterized through a $\pi$ Berry phase \cite{FractionalArticle,OneHalf}. In addition, surface states of three-dimensional topological insulators 
with one Dirac point on the upper or bottom surface are also described through a similar QAH state \cite{FuKane,SekineNomura,Zhang}. Such surfaces then give rise to a massive Dirac point with a half-quantized quantum Hall conductance
which is measured in layered systems \cite{Mogi}. We propose then to unify these two ways of thinking through the infinite alternating series related to observables such as the quantum Hall response and the response to circularly polarized light.
The geometrical analysis presented in Appendix \ref{AppendixB}, from the divergence theorem, yet shows the possibility of a $\frac{1}{2}$ topological number 
associated to the thermodynamical limit, which may also have practical applications for classical physics through the occurrence 
of a modulated electric or magnetic field alternating its sign in $z$ direction.
It is relevant to emphasize here the recent interest in relating Ramanujan infinite series with topological quantum Hall physics \cite{Montambaux}.

\subsection{Ramanujan Alternating Infinite Series: Application in Materials and $\frac{1}{2}$ Invariant at a Topological Phase Transition in Real Space through Quantum Hall Response and Light}

To acquire a physical understanding of the Ramanujan series, we can regularize the series in the sense of Abel \cite{Candelpergher}
\begin{equation}
{\mathcal S} = \hbox{lim}_{\epsilon\rightarrow 0}\left(\sum_{j=0}^{\mathcal A} (-1)^j(1-\epsilon)^j = \frac{1}{1+(1-\epsilon)} = \sum_{j=0}^{\mathcal A} (-1)^j(1-j\epsilon)\right).
\end{equation}
Suppose we apply an electric field ${\mathcal E}$ in each plane where a plane also corresponds to a sphere. 
In this way we measure the quantum Hall conductivity. 
Then, the factor $(1-\epsilon j)$ is equivalent to say that the electric field in each plane is effectively ${\mathcal E}(1-\epsilon j)$.
When $j\rightarrow +\infty$ there exists a limit for which the effective electric field is zero or equivalently the pumped transverse charge (Hall current) is zero. Equivalently, the charge remains at the north pole on each sphere associated
to each plane on the right of the interface. The charge cannot be negative  because in the view of Newton in Appendix \ref{AppendixA} when reaching the north pole a particle stays there. 
Therefore, when we meet the plane such that the quantum Hall current is zero then above this limit this is similar as if we insert an infinity of layers which are topologically trivial (e.g. an ordinary insulator of variable width) corresponding then to add an infinite number of pairs of $(1-1)$ in the Ramanujan series. This implies the presence of an interface that we can model through a topological phase transition in real space through the parameter $M$ introduced in preceding Sections: for planes below the interface each plane may be described with a Semenoff mass $M=0$ (or any $M$ smaller than $M_c$), then at the interface $M=M_c$ and above the interface the material is trivial i.e. $M>M_c$. 
In this sense, we re-interpret the Ramanujan series which is equivalent to
\begin{equation}
\label{S}
{\mathcal S} = 1-{\mathcal S} = \frac{1}{2}
\end{equation}
in a physical way as two equivalent infinite series
\begin{equation}
\label{S1}
{\mathcal S}_1 = \left(1-1+1-1 + {\color{blue} \frac{1}{2}}\right) + (1-1) + (1-1) + (1-1)+.... = \frac{1}{2}
\end{equation}
\begin{equation}
\label{S2}
{\mathcal S}_2 = \left(1-1+1-1 +1 -{\color{blue} \frac{1}{2}}\right) + (1-1) + (1-1) + (1-1)+.... = {\mathcal S}_1 = \frac{1}{2}.
\end{equation}
In the first summation, the gap closes on the layer 5 at the $K'$ Dirac point corresponding to a topological invariant $C=+\frac{1}{2}$ (mentioned in blue in the equation above) within our analysis. 
This is precisely equivalent to add a Semenoff mass on the fifth layer such that $M=M_c$. The plane 5 is similar to the first plane with $j=0$ and has a topological invariant $+1$ with the definition of the series.
The trivial region then may be thought of as planes with $M>M_c$. In the second summation, the gap closes on the layer 6 at the $K$ Dirac point corresponding then to a topological invariant $C=-\frac{1}{2}$. In this way, when saying ${\mathcal S}_1={\mathcal S}_2$ the infinity is indeed reached when we meet the interface because the situations with a even and a odd number of planes in the topological region become identical.
Multiplying  ${\mathcal S}_1={\mathcal S}_2$ by $\frac{e^2}{h}$ then this is also equivalent to measure the quantum Hall response of the material. This situation can also describe the situation where a material would have a different height
in different regions of the sample, i.e. on one region the system will be physically described through the series ${\mathcal S}_1$ and on one region the system will be physically described through the series ${\mathcal S}_2$.

Below, we show how the (physical) regularization of infinite series can also measure the response to circularly polarized light in the material. First, we evaluate the response to circularly polarized light
for the mathematical situation of an infinite number of planes with alternating topological numbers. We will then show that this response is equivalent to the materials associated to Eqs. (\ref{S1}) and (\ref{S2}).
The planes associated to an index $j$ which is even will interact with the right-handed circularly polarized light and the planes described with $j$ odd will interact with the left-handed circularly polarized light. If we measure $\kappa'$ in Eq. (\ref{kappa}) for each plane i.e. the topological invariant, then the total response for the situation of the right-handed
circularly polarized light will be
\begin{equation}
\label{serieslight}
C_{j=0} + \sum_{j=1}^{+\infty} C_{2j} \frac{1}{j^s}.
\end{equation}
The factor $\frac{1}{j^\frac{s}{2}}$ refers to the light propagation in each plane through a power-law attenuation factor for the in-coming vector potential in Sec. \ref{light}. This series can be re-summed as
\begin{equation}
\label{zeta}
C_{j=0}(1+\zeta(s)),
\end{equation}
with $\zeta(s)$ the Riemann zeta function. We suppose that each plane is characterized through a resonance with light. Assuming a power-law decay of the light signal in the bulk this would allow to reveal Eq. (\ref{zeta}), such
that the re-summation of the infinite number of planes in the neighborhood of the plane at $j=0$ will then reveal the Riemann zeta function $\zeta(s)$.
The interesting surprise then is that if we extrapolate $s\rightarrow 0$ i.e. light can reach the infinity in the vertical direction such that the result of the measure will be
\begin{equation}
C_{j=0}(1+\zeta(0)) = C_{j=0}\left(1-\frac{1}{2}B(0)\right) = \frac{1}{2}C_{j=0} = {\mathcal S}.
\end{equation}
Here, $B(0)=+1$ is the Bernoulli number. The limit $s\rightarrow 0$ may be reached when trying various forms of power-law attenuations from extrapolation analysis.
This is similar to say that compared to the result of one plane, the effect of the (many) other planes is then to compensate for an additional $-\frac{1}{2}$. This is similar to the result obtained
from the geometrical analysis in Appendix \ref{AppendixB}. We obtain the same result if we apply a left-handed circularly polarized light resonating with the planes with odd $j$ indices and measure the heights of the red peaks located at the two
Dirac points. In that case, from the definition of $\kappa'$ the result is $\frac{1}{2}C_{j=1}=-{\mathcal S}$. If the number of planes is infinite, this is equivalent as if we modify $C_{j=0}\rightarrow -C_{j=0}=C_{j=1}$ in Eq. (\ref{serieslight}).
If we measure circular dichroism in this way corresponding to measure the half difference of the two signals (i.e. of the left and right 
light polarizations) then this measures ${\mathcal S}$. 

Now, we build a correspondence with the measures of the light responses in the materials associated to the series ${\mathcal S}_1$ and ${\mathcal S}_2$.
For the material associated to the series ${\mathcal S}_1$, if we collect the information on the right-handed and left-handed circularly polarized lights, i.e. we measure $\kappa'$ in each plane through the peaks associated to the light responses, 
then two successive planes described through topological numbers $(+1 -1)$ or $(-1+1)$ will reveal zero. In this way, from the interface we will measure physically a one-half response: the light signals will reveal only one peak from the $K$ Dirac point.
For the material associated to the series ${\mathcal S}_2$, if we collect the information on the right-handed and left-handed circularly polarized lights from the planes five and six, then this will also reveal a one-half response i.e. the response
is also equivalent to the response of one Dirac point within the topological phase. This analysis then supports that the infinity is reached when meeting the interface. 

It is also important to mention that in the model of interacting spheres revealing fractional topological numbers \cite{FractionalArticle}, when taking the thermodynamical limit, we can also reach a $\frac{1}{2}$ topological number for each 
sphere. 

\section{Conclusion}
\label{summary}

We have developed the formalism of local topological invariants (markers) from the magnetic monopole to lattice models with an emphasis on the Haldane model and on its topological quantum phase transition.
The numerical analysis presented in this work through the various probes shows how this geometrical approach of local character in momentum space is efficient, simple and useful.
The local invariants are associated to physical observables such as the quantum Hall response and local response in reciprocal space to circularly polarized light. 
We introduce an effective magnetic moment for the monopole such that the susceptibility response with respect to the additional magnetic field along $z$ direction reveals the topological phase transition.
Through a proximity effect in coupled-planes systems, the QAH effect and the QSH effect are alternatively related through a even-odd effect. When taking the thermodynamical limit, we formulate a correspondence (analogy) between a $\frac{1}{2}$ invariant
at a topological interface and the Ramanujan infinite alternating series. Many challenging questions remain 
to be addressed within this geometrical approach that we hope will be useful to the community, related to practical applications and to recent results on photoluminescence results locally resolved in momentum space \cite{C2N}.
Topological properties are also measurable in circuit quantum electrodynamics lattices through a local pump probe in real space \cite{JulianKaryn}.
 \\

 K. L. H. acknowledges interesting discussions at the Conference on Quantum physics in curved spacetimes at Tours in Le Studium, Institute for Advanced Studies, Loire Valley.
A. Baldanza is thankful to Universities Paris-Saclay, Porto and Roma through the  Erasmus Mundus Quarmen Master program, and to Ecole Polytechnique for the support in his Master Thesis. 

\appendix
\section{Relation to Transport and Quantum Hall response,  Map onto the Cylinder}
\label{AppendixA}

Here, we formulate an analogy with a charge $e$ navigating from north to south in an electric field oriented along the unit vector related to the dressed polar angle. The second Newton 
equation is $\hbar \dot{\tilde{\theta}}= e{\mathcal E}$. From quantum mechanics and Parseval-Plancherel theorem, it is then possible to 
show the existence of a {\it transverse pumped current} of the form \cite{KLHReview,FractionalArticle}
\begin{equation}
J_{\perp}(\tilde{\theta}) = \frac{Q_{\perp}}{T} = \frac{e A'_{\varphi}(\tilde{\theta}=\tilde{\theta}_c^-)}{T}.
\end{equation}
The charge is measured at angle $\tilde{\theta}_c$ at time $T$ such $\hbar \tilde{\theta}_c= e {\mathcal E} T$. This form of transverse current can also be verified from many-body physics 
related to the quantum Hall response  \cite{KLHReview,FractionalArticle} and can be interpreted as a Karplus-Luttinger velocity \cite{KarplusLuttinger}.
The definition of the $A'_{\varphi}$-variable is obtained from Stokes theorem applied with two domains meeting at angle $\tilde{\theta}_c$ \cite{KLHReview,FractionalArticle}. To validate Stokes theorem and geometrical definitions for a radial magnetic field, this is equivalent to say that this requires two reference points for the vector potential \cite{Nakahara} or for the Berry gauge potential such that these two reference points may be the two poles belonging to different domains
or more generally the points defined as $\tilde{\theta}(\theta=0)$ and $\tilde{\theta}(\theta=\pi)$. On the domain related to the north pole, we can introduce the definition  \cite{KLHReview,FractionalArticle}
\begin{equation}
A'_{\varphi}(\tilde{\theta}<\tilde{\theta}_c) = A'_{\varphi}(\tilde{\theta}=\tilde{\theta}_c^-)= A_{\varphi}(\tilde{\theta}=\tilde{\theta}_c)-A_{\varphi}(\tilde{\theta}(\theta=0))
\end{equation}
and on the domain related to  $\tilde{\theta}(\theta=\pi)$ we can then introduce
\begin{equation}
A'_{\varphi}(\tilde{\theta}>\tilde{\theta}_c) = A'_{\varphi}(\tilde{\theta}=\tilde{\theta}_c^+)= A_{\varphi}(\tilde{\theta}=\tilde{\theta}_c)-A_{\varphi}(\tilde{\theta}(\theta=\pi)).
\end{equation}
These definitions can be viewed as transporting information from each pole on a thin cylinder (similar to a candle or Dirac string) on each side of the equator.
The important point is that the function $A_{\varphi}(\tilde{\theta})$ is continuous and differentiable on the whole surface and the Berry functions are then introduced with the same $\varphi$ coherent gauge, such that the topological invariant equivalent reads
\begin{equation}
C = A'_{\varphi}(\tilde{\theta}<\tilde{\theta}_c)  - A'_{\varphi}(\tilde{\theta}>\tilde{\theta}_c).
\end{equation}
In the trivial phase, $A'_{\varphi}(\tilde{\theta}<\tilde{\theta}_c)=A'_{\varphi}(\tilde{\theta}>\tilde{\theta}_c)$ which is another way to say we have only one domain.
To obtain a quantized transverse topological response within the topological phase we can e.g. measure the total pumped transverse charge when $\theta=\pi$ i.e. when $\tilde{\theta}_c=\tilde{\theta}(\theta=\pi)=\pi$. 
We measure $A'_{\varphi}(\tilde{\theta}<\tilde{\theta}_c)=C$ such that $A'_{\varphi}(\tilde{\theta}>\tilde{\theta}_c)=0$. For $M<B$ this corresponds to a navigation from north to south pole on the effective sphere associated to the dressed angle. The total transverse pumped charge then measures the topological invariant $C$. 

Since the topological invariant $C$ can be measured from the poles we can then progressively transform the geometry e.g. onto an ellipse or onto a cylinder such that information at the poles remains invariant. We assume a corresponding cylinder of fixed height $H=2$ such that the total surface area of the cylinder of unit radius i.e. $2\pi H$ is equal to the surface area on the sphere. On the top and bottom disks of the cylinder then we maintain the Berry gauge potentials to $A_{\varphi}(\tilde{\theta}(\theta=0))$ and $A_{\varphi}(\tilde{\theta}(\theta=\pi))$ respectively. The variable $u$ associated to the vertical axis of the cylinder is $u=\cos\tilde{\theta}$ such that within the topological phase $A_{\varphi}(u)=-u/2$. 
Within the topological phase, the Berry curvature is $F_{\varphi u}=-\partial_u A_{\varphi}=\frac{1}{2}$ on the whole surface area of the cylinder \cite{KLHReview} and at the transition this is equivalent to modify $F_{\varphi u}=0$ for $\tilde{\theta}>\frac{\pi}{2}$ corresponding to
maintain the Berry gauge potential fixed to $A_{\varphi}(\tilde{\theta}(\theta=\pi))=0$ in the region $\tilde{\theta}\in ]\frac{\pi}{2};\pi]$. The total transverse pumped current then is not modified in the region $\tilde{\theta}\in ]\frac{\pi}{2};\pi]$.
We can now measure the conductance located at the two edges associated with the vertical region of the surface area of the cylinder, introducing a difference of potential $H.E=(V_b-V_t)$ between bottom and top of the cylinder. 
We measure the transverse pumped current at the final time to reach $\tilde{\theta}=\pi$. 
On the cylinder geometry, the transverse pumped current then takes the form
\begin{equation}
|J_{\perp}| = \frac{e^2}{h}2EC = \frac{e^2}{h} C(V_b-V_t) = I_b-I_t.
\end{equation}
Within the topological phase with $C=1$, we identify edge currents moving in different directions $I_b$ and $I_t$ at $z=-1$ and $z=+1$ respectively satisfying the Landauer-B\" uttiker formula \cite{Buttiker,Halperin}
\begin{equation}
G = \frac{dI}{dV} = \frac{q^2}{h}C = \frac{e^2}{h}.
\end{equation}
We can then reveal the quantized conductance of an edge state. When we reach the transition $C=\frac{1}{2}$, within this formulation this corresponds to a halved current on each edge. 
For $B>M$, the transverse pumped current is zero which corresponds effectively to a charge staying at the north pole or top disk of the cylinder with $I_b-I_t=0$.

\section{Divergence Theorem for the coupled planes when reaching the infinity limit}
\label{AppendixB}

When we reach the dense limit of planes (planks), the Berry curvature takes the form
\begin{equation}
F_z = (-1)^z F_{k_x k_y} \theta(z) = e^{\pm i\pi z} F_{k_x k_y} \theta(z)
\end{equation}
with the function $F_{k_x k_y}$ which does not depend on $z$. Here, $\theta(z)$ is the Heaviside step function. This leads to
\begin{equation}
\frac{\partial F_z}{\partial z} = \pm i\pi e^{\pm i\pi z} F_{k_x k_y}\theta(z) + e^{\pm i\pi z} F_{k_x k_y} \delta(z).
\end{equation}
The divergence theorem requires us to evaluate
\begin{eqnarray}
\frac{1}{2\pi}\int_0^{+\infty} dz \iint dk_x dk_y \left(\pm i\pi e^{\pm i\pi z} F_{k_x k_y}\theta(z) + e^{\pm i \pi z} F_{k_x k_y} \delta(z)\right).
\end{eqnarray}
Integrating the second term on the variable $z$ from zero to $+\infty$ gives $\frac{C_{N=0}}{2}$. The first term is zero because we can introduce the identity $(-1)^z = \frac{e^{i\pi z} + e^{-i\pi z}}{2}$ which corresponds
to add the two terms with different signs above. From the divergence theorem, we can then interpret the term $\frac{C_{N=0}}{2}$ as the effective result on one surface when summing the effect of the infinite number of thin planes. This surface 
may be seen as the bottom plane from the $\delta(z)$ function corresponding then to slightly transport this topological charge from the interface to the first plane at $j=0$.

\bibliographystyle{crunsrt}
\bibliography{sample}

\end{document}